\documentclass[11pt]{article}

\usepackage{amsmath,amssymb,amsthm}
\usepackage{natbib}
\usepackage{enumitem}
\usepackage{xcolor}
\usepackage{tikz}
\usepackage{rotating}   
\usepackage{graphicx}   
\usepackage{adjustbox}  
\usepackage{booktabs}   
\usepackage{algorithmicx}
\usepackage{algorithm}
\usepackage{caption}
\usepackage{algpseudocode}\usepackage{soul}
\newtheorem{theorem}{Theorem}[section]

\usepackage{titlesec}
\usepackage{hhline}
\usepackage{siunitx}
\usepackage{placeins}

\def\spacingset#1{\renewcommand{\baselinestretch}{#1}\small\normalsize}

\begin{document}


\title{\bf Improving Efficiency of Regression Analyses by Integrating Data from Population-Representative Surveys: A Model-Assisted Calibration Approach}

\author{
Yanhao Lu$^1$ and Lingxiao Wang$^{1*}$\\
$^1$Department of Statistics, University of Virginia, \\Charlottesville, VA, United States\\
$^{*}$ Corresponding\ to\ lingxiao.wang@virginia.edu
}
\date{}
\maketitle

\begin{abstract} 

The increasing availability of diverse data sources has motivated great interest in data integration for improving regression efficiency. Existing data integration methods primarily focus on integrating nonprobability samples and typically assume that the integrated data sources represent the same target population. While this assumption is often difficult to justify for nonprobability samples, it is naturally satisfied when integrating probability-based surveys designed to represent a common target population. Such surveys are important research data sources because they provide representative samples and collect rich information on diverse variables, making them well suited to data integration. However, existing integration methods do not accommodate complex sampling designs. We propose model-assisted calibration methods to improve regression efficiency by integrating multiple probability-based survey samples. The proposed framework accommodates settings in which either individual-level data or only summary statistics are available from external surveys while preserving valid finite-population inference without requiring correct specification of the outcome model. We establish the design consistency of the proposed estimators and develop Taylor linearization variance estimators accounting for the complex sampling designs of both surveys. Simulation studies and an application integrating National Health and Nutrition Examination Survey and National Health Interview Survey demonstrate substantial efficiency gains while maintaining valid finite-population inference.
\end{abstract}


\newpage
\spacingset{1.45}   

\section{Introduction}

Integrating information across multiple data sources to improve efficiency of regression analyses has become increasingly important in modern statistical practice. In many applications, particularly in medical studies and health surveys, certain variables can be costly to measure and are therefore observed only in a relatively small \lq\lq internal\rq\rq\ study sample. In contrast, inexpensive covariates and some other auxiliary information are often available in a larger \lq\lq external\rq\rq\ sample, such as large-scale cohort studies, questionnaire-based national health surveys, administrative databases, and ongoing public health surveillance systems. Borrowing information from these large external data sources can substantially improve the efficiency of estimation without incurring additional data collection costs.

Statistical approaches have been proposed to improve the efficiency of regression analysis by borrowing strength of external data sources using their individual-level  auxiliary information (microdata) \citep{ angelopoulos2023prediction} or summary statistics when the microdata is not accessible due to confidentiality or data-sharing restrictions\citep{chatterjee2016constrained, zhang2020generalized,zheng2022risk}.
Most of these approaches assume that the internal sample and the external sample represent the same target population or the outcome model is correctly specified. These two assumptions can be easily violated if one or both of the internal and the external studies recruit eligible volunteers that have heterogeneous distributions of covariates and even the outcome. As a result, directly borrowing the summary statistics from the external sample can introduce bias in estimating regression coefficients \citep{wang2025using}.

On the contrary, the assumption of a homogeneous population is naturally held if both internal and external data are collected from population-representative surveys of the same target population. For example, both the US National Health and Nutrition Examination Survey (NHANES) and the US National Health Interview Survey (NHIS) are repeated cross-sectional studies that annually select representative random samples of individuals from a target finite population (FP) of non-institutionalized US residents by stratified multistage cluster sampling designs. In the first stage, primary sampling units (PSUs) consisting of (groups of contiguous) counties (or cities and parts of large cities) are selected with probability proportional to measures of size (PPS) within each stratum defined by geography (e.g., census regions). In later stages, units consisting of segments of census blocks, households, and individuals are randomly selected using known selection probabilities within each unit sampled in the previous stage. The resulting samples, with the complex sampling designs considered, can well represent the target FP and provide design-consistent estimators of the FP quantities such as regression coefficients without relying on the correct specification of outcome models. 

Due to different interview modes (e.g., mail, telephone, or face-to-face survey), and types of collected variables, the expected sample sizes of these surveys can be substantially different. For example, NHIS, as the nation\rq s largest and oldest national health survey in the US, yields approximately 27,000 adult interviews and 9,000 children interviews annually. It collects a broad range of variables via face-to-face interviews, including demographics and socioeconomic characteristics, health status (e.g., chronic conditions, disabilities, and mental health), health behaviors (e.g., diet, exercise, and smoking), healthcare access, and family/household composition. NHANES, besides the self-reported variables collected from face-to-face interviews, also conducts physical examinations to obtain medical measurements and results of laboratory tests from biospecimens such as blood and urine. Therefore, NHANES provides especially richer data for nationally representative medical studies than NHIS. However, due to the high costs of the physical exams and the biospecimens collection and measurements, NHANES only samples about 5,000 individuals annually (about 10,000 in each 2-year survey cycle). Some variables are measured only in a random subset of NHANES participants, leading to even smaller sample sizes and consequently limiting the efficiency of the resulting regression analyses. Integrating information collected from NHIS can be particularly beneficial for improving the efficiency of regression analyses based solely on NHANES data.

One of the major challenges in integrating data from multiple national surveys is appropriately accounting for their stratified multistage cluster sampling designs, which cannot be easily accommodated by existing likelihood-based approaches \citep{zhang2020generalized}. In contrast, survey weight calibration provides a natural design-based framework for incorporating auxiliary information while accounting for complex survey designs. It is widely used to improve survey estimators of population totals and means by incorporating summary statistics of auxiliary variables obtained from the target population or larger external surveys \citep{deville1992calibration, wu2001model,sarndal2007calibration}. In more recent literature, weight calibration has been extended to regression analyses under another data structure of two-phase medical studies \citep{lumley2011,shin2020weight}  and two-phase health surveys \citep{wang2025using} 
where the small Phase II sample measuring additional variables is a random subset of the large Phase I sample. By borrowing auxiliary information from the large Phase I sample, calibration methods (that include generalized raking) can substantially improve the efficiency of Phase II regression analyses while remaining robust to model misspecification, because they avoid the strong distributional assumptions required by likelihood-based or imputation approaches. However, these existing calibration methods rely on individual-level Phase I microdata and the assumption that the Phase II sample is a random subset of the Phase I sample. Consequently, they cannot be readily extended to general survey integration settings, especially when only summary statistics rather than microdata are available from the external survey.

In this paper, we propose model-assisted calibration methods to improve regression estimation by integrating data from two population-representative surveys. Under the design-based sampling framework, the proposed methods do not rely on correct specification of the outcome model, making them robust to model misspecification. They accommodate settings in which either individual-level microdata or only summary statistics are available from the external survey. We establish the design consistency of the proposed calibration estimators and develop Taylor linearization variance estimators that properly account for the complex sampling designs of both the internal and external surveys.

The remainder of this paper is organized as follows. Section 2 introduces the regression model of interest, the notation, the data structure, and the underlying assumptions. Sections 3 and 4 present the proposed calibration methods for settings with and without access to microdata from the external survey, respectively. Section 5 establishes the statistical properties of the proposed estimators and derives Taylor linearization variance estimators. Section 6 evaluates the finite-sample performance of the proposed methods in terms of efficiency and robustness through a series of simulation studies. We then illustrate the practical utility of the proposed methods by estimating the association between all-cause mortality and total body fat through the integration of NHANES and NHIS data in section 7. Finally, Section 8 concludes with a discussion of the implications, limitations, and directions for future research.


\section{Model and Notation}

We are interested in estimating the association between the outcome $Y$ and a set of covariates $\pmb{X} = (\pmb{X}_1^\top, X_2)^\top \in \mathbb{R}^{p}$ in a finite population (FP) using the following working generalized linear model
\begin{align}
g\{ E(Y \mid \pmb{X}) \} = \pmb{\beta}_1^\top \pmb{X}_{1} + \beta_2 X_2, 
\label{eq:logit}
\end{align}
where $\pmb{X}_1$ is a set of easy-to-obtain covariates and $X_2$ is a covariate that is costly to measure, \( g(\cdot) \) is a known link function depending on the type of outcome \( Y \), and \( \pmb{\beta} = (\pmb{\beta}_1^\top, \beta_2)^\top \) is the vector of regression coefficients. Without loss of generality, we assume that the intercept is included in $\pmb{X}_1$. For simplicity of notation, we present the methodology
using one expensive covariate $X_2$, although the proposed framework naturally
extends to multiple expensive covariates (and interactions) or to expensive outcome $Y$.

Let the set of individuals in the FP be denoted by \( \mathcal{U} = \{1, \ldots, N\} \) where the values of $Y$ and $\pmb{X}$ are denoted by $\{(y_i, \pmb{x}_i),i\in \mathcal{U}\}$. If $\{(y_i, \pmb{x}_i),i\in \mathcal{U}\}$ are fully observed, then $\pmb{\beta}$ in the outcome model \eqref{eq:logit} can be estimated by solving the FP estimating equation 
\begin{align}
\pmb{U}(\pmb{\beta}) = \sum_{i \in \mathcal{U}} \pmb{u}_{i}(\pmb{\beta}) = \pmb{0}, 
\label{eq:fp_score}
\end{align}
where $\pmb{u}_i(\pmb{\beta})$ is the contribution of individual $i$ to the score function $\pmb{U}(\pmb{\beta})$. For example, $\pmb{u}_i(\pmb{\beta}) = \{y_i - g^{-1}(\pmb{\beta}^\top \pmb{x}_i)\} \pmb{x}_i$ when $g(\cdot)$ is the identity, logit, or log link, for continuous, binary, and count outcome $Y$ respectively. We use $\pmb{\beta}_{FP}$ to denote the solution for $\pmb{\beta}$ in Equation (\ref{eq:fp_score}), which is the regression coefficients in the FP. Besides $Y$ and $\pmb{X}$, we consider a set of ancillary variables $\pmb{Z}$ with value of $\pmb{z}_i$ for $i\in \mathcal{U}$
that may be correlated with the expensive covariate $X_2$. We assume that $\pmb{Z}$ are not associated with  with $Y$ conditional on $\pmb{X}$ and therefore are not included in the outcome model (\ref{eq:logit}). Note that the association between
$\pmb{Z}$ and $X_2$ can help with the efficiency gain of the proposed estimator of $\pmb{\beta}$, but is not required for its consistency.

\FloatBarrier

\begin{figure}[ht]
\centering
\includegraphics[width=\textwidth]{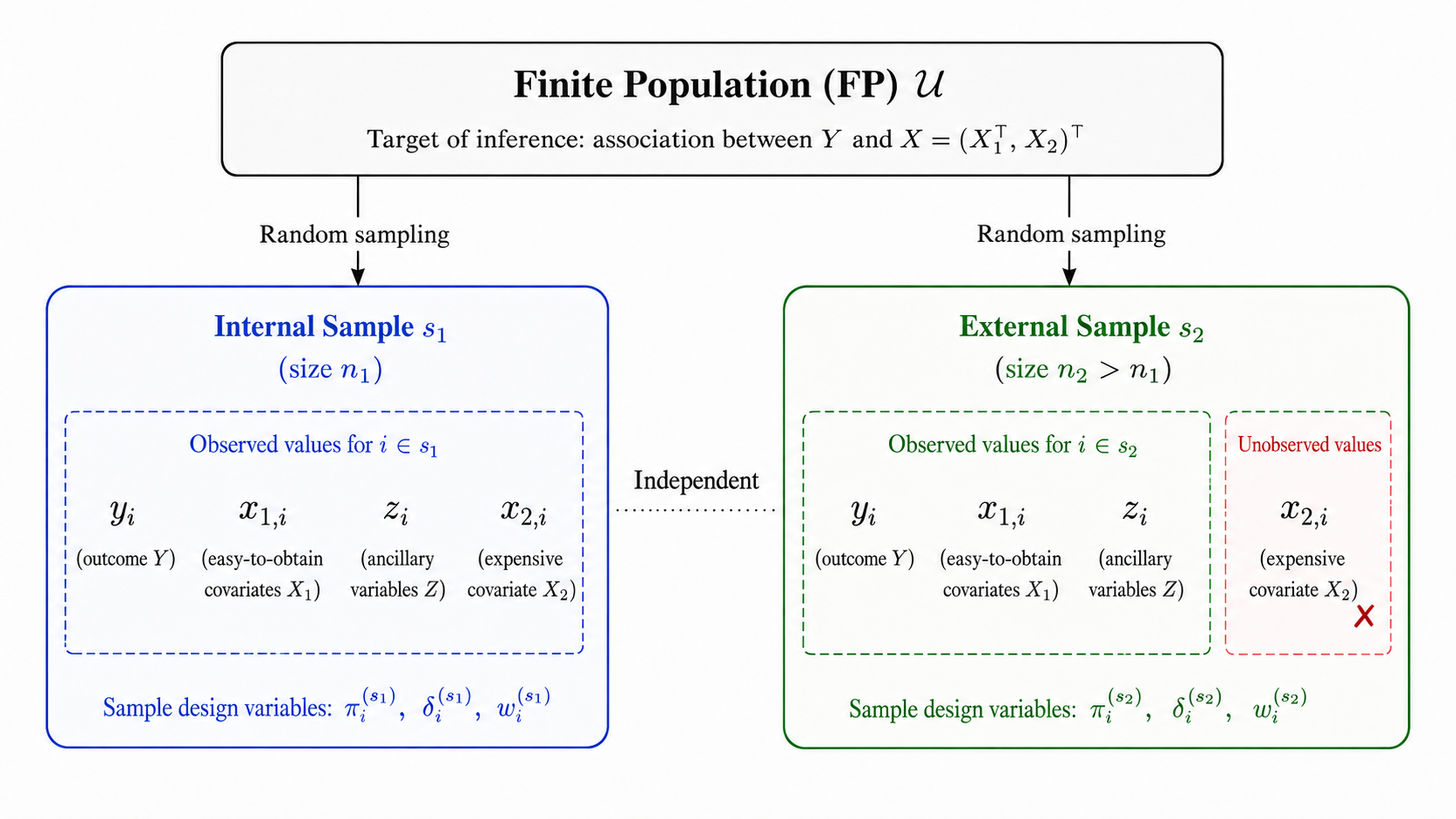}
\vskip-1cm
\caption{\small
Structure of observed and unobserved data in the internal sample $s_1$ and the external sample $s_2$.
}
\label{fig:data_structure}
\end{figure}

\FloatBarrier
\noindent In practice, however, \( (y_i, \pmb{x}_i) \) is not available for the entire $\mathcal{U}$. We consider the structure of the observed and unobserved data in Figure~\ref{fig:data_structure}, with the details described as follows.


The \textit{internal sample} \( s_1 \) of size \( n_1 \) is randomly selected from the FP with inclusion probabilities $\{\pi_i^{(s_1)}, i \in \mathcal{U}\}$, which are known from the sampling design. We use $\delta_i^{(s_1)}$ to indicate whether $i \in \mathcal{U}$ is selected into $s_1$ (i.e., $\delta_i^{(s_1)} = 1$ for $i \in s_1$, and $\delta_i^{(s_1)} = 0$ for $i \in \mathcal{U} \setminus s_1$), with $P(\delta_i^{(s_1)} = 1) = \pi^{(s_1)}_i$. We use $w_i^{(s_1)} = 1 / \pi_i^{(s_1)}$ to denote the corresponding inverse inclusion probability sampling weight. We observe $ \{(y_i, \pmb{x}_{1,i}, x_{2,i}, \pmb{z}_{i}, w^{(s_1)}_i), i \in s_1\}$ in the internal sample $s_1$. Therefore, $\pmb{\beta}_{FP}$ can be approximately unbiasedly estimated by solving the $s_1$-sample-weighted estimating equation $\widehat{\pmb{U}}^{(s_1)}(\pmb{\beta}) = \sum_{i \in s_1} w_i^{(s_1)} \pmb{u}_i(\pmb{\beta}) = \pmb{0}$. Let $\widehat{\pmb{\beta}}_{s_1}$ be the corresponding $s_1$ sample-weighted estimator of $\pmb{\beta}_{FP}$.

The \textit{external sample} \( s_2 \) is also a random sample of $\mathcal{U}$, with the sample size $n_2 > n_1$. Analogous to the notation for sampling design variables of $s_1$, we use $\pi_i^{(s_2)}$, $\delta_i^{(s_2)}$, and $w_i^{(s_2)}$ to denote the inclusion probabilities, sampling indicators, and the corresponding sample weights, respectively. In \( s_2 \), we observe \( \{(y_i, \pmb{x}_{1,i}, \pmb{z}_{i}, w_i^{(s_2)}), i \in s_2\} \), but not the values of $X_2$.  

We make the following standard assumptions for $s_1$ and $s_2$:
\begin{description}
  \item[(A.1)] \( s_1 \) and \( s_2 \) are randomly selected from the same FP with positive inclusion probabilities, i.e., $0 < \pi_i^{(k)} \le 1$, $k = 1, 2$.\\[-30pt]
  \item[(A.2)] $s_1$ and $s_2$ are independently selected, i.e., $\mathrm{Cov}(\delta^{(s_1)}_i, \delta^{(s_2)}_j) = 0$ for $i, j \in \mathcal{U}$.

\end{description}
Assumption (A.1) along with additional regularity conditions in Supplementary Materials Section 1.1  ensures that the FP quantities can be consistently estimated using $s_1$, $s_2$. Assumption (A.2) simplifies the variance of the estimated regression parameters using the integrated data from $s_1$ and $s_2$. Besides these two fundamental assumptions (A.1) and (A.2), we require that variables $Y$, $\pmb{X}_1$, $\pmb{Z}$ are measured in the same way in $s_1$ and $s_2$ to avoid bias due to measurement errors or inappropriate data harmonization.. 


Under Assumption (A.1), $\widehat{\pmb{\beta}}_{s_1}$, the $s_1$ estimator of $\pmb{\beta}_{FP}$, is design consistent, but its variance can be large due to the small sample size. In order to seek a way for reducing the variance of $\widehat{\pmb{\beta}}_{s_1}$, we show that 
\begin{align}\mathrm{Var}(\widehat{\pmb{\beta}}_{s_1}) = \mathrm{Var}\left\{ \sum_{i \in \mathcal{U}} \delta_i^{(s_1)} w_i^{(s_1)} \Delta_i \right\}+o(n_1^{-1}),
\label{eq:s1.var}
\end{align}
where $\mathrm{Var}(\cdot)$ is the sampling variance with respect to the sampling indicator $\delta_i^{(s_1)} $, and $\Delta_i=\left.\left\{\pmb{U}^{-1}_{\pmb{\beta}}\pmb{u}_i(\pmb{\beta})\right\}\right| _{\pmb{\beta}=\pmb{\beta}_{FP}}$ is the influence function of $\pmb{\beta}$, with $\pmb{U}_{\pmb{\beta}}=E\{\partial\pmb{U}(\pmb{\beta})/\partial\pmb{\beta}\}$, where $E\{\cdot\}$ is with respect to   $\delta_i^{(s_1)}$.
Formula (\ref{eq:s1.var}) shows that high variability of $\{w_i^{(s_1)} \Delta_i,i\in s_1\}$ may lead to a large sampling variance of $\widehat{\pmb{\beta}}_{s_1}$. Therefore, to reduce the variance of $\widehat{\pmb{\beta}}_{s_1}$, we propose calibration methods to assign a weight adjustment factor $F_i$ for $i\in s_1$ using information from $s_2$ so that $\{F_iw_i^{(s_1)}  \Delta_i,i\in s_1\}$ have less variability, leading to  estimators of $\pmb{\beta}_{FP}$ that are more efficient than $\widehat{\pmb\beta}_{s_1}$. We consider two approaches for calculating adjustment factors $F_i$. The first approach utilizes microdata from $s_2$ (Section 3) while the second approach utilizes only summary statistics from $s_2$ (Section 4).

\section{Pooled-Sample Weight Calibration with Micro Data of $s_2$}


With access to microdata from $s_2$, we can pool $s_1$ and $s_2$ to obtain a larger study sample, denoted by $S = s_1 \cup s_2$ with sample size $n = n_1 + n_2$ and observed data $\{(y_i, \pmb{x}_{1,i}, \pmb{z}_i), i \in S\}$. The internal study sample is treated as a subset of the combined sample, i.e., $s_1 \subset S$, with additional observations $\{x_{2,i},i\in s_1\}$ for $X_2$ in $s_1$. 

This data structure is similar to two-phase study designs, where the outcome $Y$ and easy-to-obtain covariates $\pmb{X}_1$ are usually available in the entire study sample (Phase I sample, analogous to $S$), and the expensive predictor $X_2$ is exclusively measured in a random subset (Phase II sample) of the Phase I sample (analogous to $s_1$). However, the major difference is that $s_1$ and $s_2 = S \setminus s_1$ under a two-phase study design are two complementary subsets of the same study sample $S$, whereas under our data integration setting, $s_1$ and $s_2$ are independent random samples of the FP.



To address this difference, we adapt the weight calibration approach to the data integration setting by calibrating the sample weighted $s_1$ to the sample weighted $S$ using auxiliary variables that are correlated with the influence function of $\widehat{\pmb\beta}_{s_1}$  and are available in the entire pooled study sample $S$. Different from the weight calibration in traditional two-phase studies, we account for the sample weights of both $s_1$ and $s_2$ in constructing the calibration auxiliary variables, as well as in calculating the calibration adjustment factor $F_i$. The procedure for obtaining the efficient pooled-sample calibration estimator of $\pmb{\beta}$, denoted by $\widehat{\pmb\beta}_\text{p.clb}$, is described in Algorithm~\ref{alg:indiv-calib}. 


\noindent\rule{\linewidth}{1pt}
\vskip-0.3cm
\begingroup
\makeatletter
\algrenewcommand\alglinenumber[1]{\footnotesize\alph{ALG@line}:}
\makeatother
\captionof{algorithm}{Pooled-Sample Weight Calibration with Microdata from $s_2$}
\label{alg:indiv-calib}
\vskip-0.3cm
\noindent\rule{\linewidth}{0.5pt}
\begin{algorithmic}[1]
\State \textbf{Input:} $\{(y_i, \pmb{x}_i, \pmb{z}_i, w^{(s_1)}_i), i \in s_1\}$, $\{(y_i, \pmb{x}_{1,i}, \pmb{z}_i, w^{(s_2)}_i), i \in s_2\}$.

\State Using $s_1$, fit a prediction model $f(\pmb{x}_{1,i}, \pmb{z}_i)$ for $x_{2,i}$ and define
$x^{*}_{2,i}=f(\pmb{x}_{1,i}, \pmb{z}_i)$ as the predicted value of $x_{2,i}$ for $i\in S$.

\State Generate the sample weights for $S$ by $w_i^{(S)} = a_k \cdot w_i^{(s_k)}$ for $i \in s_k$, where $a_k = n_k / (n_1 + n_2)$, $k = 1,2$.

\State Fit a surrogate outcome model to the weighted sample $S$ that uses $x^*_{2,i}$ to replace $x_{2,i}$ in the outcome model \eqref{eq:logit}, and obtain the estimator of the surrogate model parameters $\widehat{\pmb{\beta}}^*_S$ by solving
\[
\widehat{\pmb{U}}^{*(S)}(\pmb{\beta}^*) = \sum_{i \in S} w_i^{(S)} \pmb{u}_i^*(\pmb{\beta}^*) = \pmb{0},
\]
where $\pmb{u}^*_i(\pmb{\beta}^*) = (y_i - p_i^*) \pmb{x}_i^*$,  $p_i^* = \mathrm{expit}\left(\pmb{x}_i^{*\top} \pmb{\beta}^*\right)$, and $\pmb{x}_i^{*}=(\pmb{x}_{1,i}^\top, x^*_{2,i})^\top$.

\State Obtain the sample plug-in influence functions of $\widehat{\pmb{\beta}}^*_S$
\[
\widehat{\Delta}_i^*
=
\left.\left\{\widehat{\pmb{U}}_{\pmb{\beta}^*}^{*(S)-1} \pmb{u}_i^*(\pmb{\beta}^*)\right\}\right|_{\pmb{\beta}^* = \widehat{\pmb{\beta}}^*_S},
\quad \text{where } \widehat{\pmb{U}}^{*(S)}_{\pmb{\beta}^*}
= \frac{\partial \widehat{\pmb{U}}^{*(S)}(\pmb{\beta}^*)}{\partial \pmb{\beta}^*}.
\]

\State Calibrate the weighted sample $s_1$ to the weighted sample $S$ using influence functions such that
\begin{align}
\sum_{i \in s_1} w_i^{(s_1)} F_i(\pmb{\eta}) \widehat{\Delta}_i^*
=
\sum_{i \in S} w_i^{(S)} \widehat{\Delta}_i^*,
\label{eq:calibfun}
\end{align}
where $F_i(\pmb{\eta})$ is the calibration adjustment factor, and $\pmb{\eta}$ is a nuisance parameter, with the solution of \eqref{eq:calibfun} denoted by $\widehat{\pmb{\eta}}$.

\State Generate the pooled-sample calibration weights $w_i^{(\mathrm{p.clb})} = w_i^{(s_1)} F_i(\widehat{\pmb{\eta}})$, and obtain the calibration estimator $\widehat{\pmb{\beta}}_{\mathrm{p.clb}}$ by solving
\[
\widehat{\pmb{U}}^{\mathrm{p.clb}}(\pmb{\beta}) = \sum_{i \in s_1} w_i^{(\mathrm{p.clb})} \pmb{u}_i(\pmb{\beta}) = \pmb{0}.
\]
\end{algorithmic}
\endgroup
\noindent\rule{\linewidth}{1pt}
\noindent The weight adjustment factors $\{F_i(\pmb{\eta}),i\in s_1\}$ depend on a user-chosen distance function that measures the distance between $\{ w_i^{(s_1)} F_i(\pmb{\eta}),i\in s_1\}$ and $\{ w_i^{(s_1)},i\in s_1\}$. Different distance measures can lead to different design-consistent estimators $\widehat{\pmb{\beta}}_{\mathrm{p.clb}}$. For example, $F_i (\pmb{\eta})= \exp\{\pmb{\eta}^\top \widehat{\Delta}_i^*\}$ provides a generalized raking estimator (GR)  \citep{lumley2011}, and $F_i (\pmb{\eta})= (1 + \pmb{\eta}^\top \widehat{\Delta}_i^*)$ provides a generalized regression estimator (GREG) , which can be approximately written as an augmented estimator as shown in Theorem~\ref{thm:consistency} below.

\begin{theorem}[Consistency of the Calibration Estimator]\label{thm:consistency}
The GREG-type pooled-sample calibration estimator $\widehat{\pmb{\beta}}_{\mathrm{p.clb}}$ can be approximately written as the following augmented estimator:
\begin{equation}
\widehat{\pmb{\beta}}_{\mathrm{p.clb}}
=
\widehat{\pmb{\beta}}_{s_1}
+
\pmb{\Theta}^{\top}
\left(
\widehat{\pmb{\beta}}^{*}_S
-
\widehat{\pmb{\beta}}^{*}_{s_1}
\right)
+
o_p(n_1^{-1/2}).
\label{eq:p.clb-theta}
\end{equation}
\end{theorem}

\noindent where  
\begin{align}
\pmb{\Theta}
=
\left(
\sum_{i \in s_1}
w_i^{(s_1)}
\Delta_i^{*\top}\Delta_i^*
\right)^{-1}
\left(
\sum_{i \in s_1}
w_i^{(s_1)}
\Delta_i^{*\top}\Delta_i
\right),
\label{eq:theta-def}
\end{align} 
which estimates the correlation between $\widehat{\pmb{\beta}}_{s_1}$ and $\widehat{\pmb{\beta}}_{s_1}^*$ using the unknown FP quantities $\{(\Delta_i, \Delta_i^*), i\in FP\}$, with $\Delta^*_i=\left.\pmb{U}_{\pmb{\beta}^*}^{*-1}\pmb{u}_i^*(\pmb{\beta}^{*})\right|_{\pmb{\beta}^*=\pmb{\beta}^{*}_{FP}}$ and $\pmb{U}^*_{\pmb{\beta}^*}=E\left(\widehat{\pmb{U}}^{*(S)}_{\pmb{\beta}^*}\right)$. We use $\widehat{\pmb{\beta}}_{\mathrm{p}}(\pmb{\Theta}):=\widehat{\pmb{\beta}}_{s_1}
+
\pmb{\Theta}^{\top}
\left(
\widehat{\pmb{\beta}}^{*}_S
-
\widehat{\pmb{\beta}}^{*}_{s_1}
\right)$ to denote the augmented estimator on the right side of formula (\ref{eq:p.clb-theta}). 

Under Assumptions (A.1)–(A.2) and standard regularity conditions in Supplementary Materials Section 1.1, we have  
\begin{align}
\widehat{\pmb{\beta}}_{\mathrm{p.clb}} \stackrel{d}{\rightarrow} 
N\left(\pmb{\beta}_{FP}, \ \mathrm{Var}(\widehat{\pmb{\beta}}_{\mathrm{p.clb}})\right),
\end{align}
where $\mathrm{Var}(\cdot)$ is the sampling variance with respect to the random sampling indicators $\delta^{(s_1)}$ and $\delta^{(s_2)}$. The closed form of $\mathrm{Var}(\widehat{\pmb{\beta}}_{\mathrm{p.clb}})$ and its approximately unbiased estimator are described in Section~5. A detailed derivation is provided in Supplementary Materials Section 1.5.

\textit{Remark}. Under Assumption (A.1), since the first summand of $\widehat{\pmb{\beta}}_{\mathrm{p.clb}}$, i.e., $\widehat{\pmb{\beta}}_{s_1}$, is design-consistent for $\pmb{\beta}_{FP}$, and the second summand is consistent for $\pmb{0}$, $\widehat{\pmb{\beta}}_{\mathrm{p.clb}}$ is design consistent for $\pmb{\beta}_{FP}$. The efficiency of $\widehat{\pmb{\beta}}_{\mathrm{p.clb}}$ depends on the correlation between $\widehat{\pmb{\beta}}^*_{s_1}$ and $\widehat{\pmb{\beta}}_{s_1}$ in  $s_1$ measured by $\pmb{\Theta}$, and can be affected by how well $x_2^*$ predicts $x_2$. When $x_2^*$ is highly correlated with $x_2$, $\pmb{\Theta}$ will be nearly equal to the identity matrix $\pmb{I}$, and therefore $\widehat{\pmb{\beta}}_{\mathrm{p.clb}}$ will be close the  estimator of $\pmb{\beta}_{FP}$ using the pooled sample $S$ if $x_{2,i}$ is observed in $s_2$, which is  more efficient than $\widehat{\pmb{\beta}}_{s_1}$ due to the larger sample size of $S$. When $x_2^*$ is not correlated with $x_2$, the calibration estimator of $\beta_2$ will be close to the original $s_1$ estimator, but the calibration estimator for $\pmb{\beta}_1$ (the regression coefficients for inexpensive covariates $\pmb{X}_1$) can still achieve substantial efficiency gains.

We also note that $\widehat{\pmb{\beta}}_{\mathrm{p.clb}}$ is nearly as efficient as the augmented estimator $\widehat{\pmb{\beta}}_{\mathrm{p}}(\pmb{\Theta})$, which requires the unknown FP influence functions $\Delta_i$ and $\Delta_i^*$ for $i \in s_1$ to construct $\pmb{\Theta}$, which cannot be substituted by the sample plug-in quantity $\widehat{\pmb{\Theta}}=
\{\sum_{i\in s_1}w_i^{(s_1)}\widehat{\Delta}_i^*\widehat{\Delta}_i^*
\}^{-1}
\{\sum_{i\in s_1}w_i^{(s_1)}\widehat{\Delta}_i^*\widehat{\Delta}_i
\}$, where $\widehat{\Delta}_i^*$ (defined in step e of Algorithm 1) and
$\widehat{\Delta}_i
= 
\{
\widehat{\pmb{U}}_{\pmb{\beta}}^{(s_1)-1} \pmb{u}_i(\pmb{\beta}) 
\} 
|_{\pmb{\beta}=\widehat{\pmb{\beta}}_{s_1}}$ are sample plug-in influence functions. This is because $\widehat{\Delta}_i$ and $\widehat{\Delta}_i^*$ use the estimated Hessian matrices $\widehat{\pmb{U}}_{\pmb{\beta}}^{(s_1)}$ and $\widehat{\pmb{U}}_{\pmb{\beta}}^{*(S)}$ respectively, which introduce non-ignorable remainders in $\widehat{\pmb{\Theta}}$ and can cause bias as well as efficiency loss of the resulting augmented estimator denoted by $\widehat{\pmb{\beta}}_{\mathrm{p}}(\widehat{\pmb{\Theta}})$ (see Supplementary Materials, Section 1.10). We provide an empirical comparisons among $\widehat{\pmb{\beta}}_{\mathrm{p.clb}}$,  $\widehat{\pmb{\beta}}_{\mathrm{p}}(\pmb{\Theta})$ and $\widehat{\pmb{\beta}}_{\mathrm{p}}(\widehat{\pmb{\Theta}})$ in Section~6 to support the above statements about their (un)biasedness and efficiencies.

\section{External-Sample Weight Calibration with Summary Data of $s_2$}

In many cases, the availability of microdata from the external sample $s_2$ may be inaccessible due to reasons such as data confidentiality, restricted data-use agreements, or limited computational resources. Accessing to some variables collected by national health surveys, such as cause-specific mortality, geographic information on participants' residence, and certain variables obtained from biosamples  (e.g., Genotyping, microbiome), requires special authorization and secure data environments at Restricted Data Centers (RDCs). These constraints motivate the idea of performing calibration using only \emph{summary statistics} from the external data, such as regression coefficient estimates and their associated variance–covariance matrices obtained from RDCs or previously published studies, rather than microdata. 

An important implication of Theorem~\ref{thm:consistency} is that the proposed GREG-type calibration estimator can be written as an augmented estimator $\widehat{\pmb{\beta}}_{\mathrm{p}}(\pmb{\Theta})$, which only requires $\widehat{\pmb{\beta}}^{*}_{S}$ from $S$ and microdata from $s_1$. When microdata from $s_2$ are not available to be pooled with $s_1$, it is natural to consider the following external-sample augmented estimator 
\begin{equation}
\widehat{\pmb{\beta}}_{\mathrm{e}}(\pmb{\Theta}) = \widehat{\pmb{\beta}}_{s_1}
      + \pmb{\Theta}^{\top}
        \left(
          \widehat{\pmb{\beta}}^{*}_{s_2} - \widehat{\pmb{\beta}}^{*}_{s_1}
        \right),
\label{eq:e.clb-theta}
\end{equation}
which substitutes the $s_2$ estimator $\widehat{\pmb{\beta}}^{*}_{s_2}$ for $\widehat{\pmb{\beta}}^{*}_{S}$ in the pooled-sample augmented estimator $\widehat{\pmb{\beta}}_{\mathrm{p}}(\pmb{\Theta})$. However, as $\widehat{\pmb{\beta}}_{\mathrm{e}}(\pmb{\Theta})$ cannot be directly obtained in practice because $\pmb{\Theta}$ is a function of unknown FP influence functions, we propose an external-sample calibration estimator $\widehat{\pmb{\beta}}_{\mathrm{e.clb}}$ that only uses $\widehat{\pmb{\beta}}^*_{s_2}$ together with the microdata from $s_1$. In Section 6, simulations show that $\widehat{\pmb{\beta}}_{\mathrm{e.clb}}$ is often nearly as efficient as $\widehat{\pmb{\beta}}_{\mathrm{e}}(\pmb{\Theta})$.

\newtheorem{lemma}[theorem]{Lemma}
\begin{lemma}[Equivalence of calibration using influence functions and score functions as auxiliary variables under generalized linear models]
\label{lem:score-if-equivalence}
$\widehat{\pmb{\beta}}_{\mathrm{e.clb}} = \widehat{\pmb{\beta}}_{\mathrm{e}}(\pmb{\Theta})+o_p(n^{-1/2})$ in equation \eqref{eq:e.clb-theta}
can be obtained by calibrating the sample-weighted $s_1$ to the sample-weighted $s_2$ using the score contributions $\pmb{u}_i^{*}(\widehat{\pmb{\beta}}^*_{s_2})$ for $i \in s_1, s_2$, which is algebraically equivalent to calibration using the (unavailable) $s_2$ sample influence functions $\widehat{\Delta}_i^*(\widehat{\pmb{\beta}}^*_{s_2})$. (Proof in Supplementary Materials Section 1.6).
\end{lemma}

The procedure for obtaining $\widehat{\pmb{\beta}}_{\mathrm{e.clb}}$ is described below in Algorithm~\ref{alg:summary-calib}.

\noindent\rule{\linewidth}{1pt}
\vskip-0.3cm
\begingroup
\makeatletter
\algrenewcommand\alglinenumber[1]{\footnotesize\alph{ALG@line}:}
\makeatother
\captionof{algorithm}{External-sample calibration with summary data from $s_2$}
\label{alg:summary-calib}
\vskip-0.3cm
\noindent\rule{\linewidth}{0.5pt}
\begin{algorithmic}[1]
\State \textbf{Input:} $\{(y_i, \pmb{x}_i^*, x_{2,i}, w^{(s_1)}_i), i \in s_1\}$, and $\widehat{\pmb{\beta}}^*_{s_2}$

\State Construct the calibration auxiliary variables using the score contributions $\pmb{u}_i^{*}(\widehat{\pmb{\beta}}^*_{s_2})$ for $i \in s_1$ (e.g., $\pmb{u}_i^{*}(\widehat{\pmb{\beta}}^*_{s_2}) = (y_i - \widehat{p}_{s_2,i}^{*}) \pmb{x}_i^{*}$, with $\widehat{p}_{s_2,i}^{*} = \mathrm{expit}(\pmb{x}_i^{*\top} \widehat{\pmb{\beta}}^*_{s_2})$ under a logit link function for binary outcome $y_i$).

\State Calibrate the sample-weighted $s_1$ to the sample-weighted $s_2$ using $\pmb{u}_i^{*}(\widehat{\pmb{\beta}}^*_{s_2})$ by solving
\begin{equation}
\sum_{i \in s_1} w_i^{(s_1)} F_i(\pmb{\eta}) \pmb{u}_i^*(\widehat{\pmb{\beta}}^*_{s_2}) = \pmb{0},
\label{eq:Qeta-simple}
\end{equation}
where $F_i(\pmb{\eta}) = 1 + \pmb{\eta}^\top \pmb{u}_i^*(\widehat{\pmb{\beta}}^*_{s_2})$, and the solution of \eqref{eq:Qeta-simple} is denoted by $\widehat{\pmb{\eta}}$.

\State Define the external-sample calibration weights $w_i^{(\mathrm{e.clb})} = w_i^{(s_1)} F_i(\widehat{\pmb{\eta}})$ and solve
\[
\sum_{i \in s_1} w_i^{(\mathrm{e.clb})} \pmb{u}_i(\pmb{\beta}) = \pmb{0}
\]
to obtain the estimator $\widehat{\pmb{\beta}}_{\mathrm{e.clb}}$.
\end{algorithmic}
\endgroup
\noindent\rule{\linewidth}{1pt}

\begin{theorem}[Consistency of the External-Sample Calibration Estimator]\label{thm:econsistency}
Under Assumptions~(A.1)--(A.2) and standard regularity conditions (see Supplementary Materials, Section~1.1),
\begin{align}
\widehat{\pmb{\beta}}_{\mathrm{e.clb}} \stackrel{d}{\rightarrow}
N\!\left(\pmb{\beta}_{FP},\ \mathrm{Var}(\widehat{\pmb{\beta}}_{\mathrm{e.clb}})\right).
\end{align}
\end{theorem}

\noindent The detailed proof is provided in Supplementary Materials Section~1.9. The closed form of $\mathrm{Var}(\widehat{\pmb{\beta}}_{\mathrm{e.clb}})$ and its approximately unbiased estimator (without requiring microdata from $s_2$) are provided in Section~5.

\section{Sampling Variance and Variance Estimation}

We derive the FP variances of the calibrated estimators  \( \widehat{\pmb{\beta}}_{\mathrm{p.clb}} \) and \( \widehat{\pmb{\beta}}_{\mathrm{e.clb}} \)  that account for the variability of the sampling of both \(s_1\) and \(s_2\) and provide the Taylor linearization variance estimates. Let \( \pmb{\theta} = (\pmb{\beta}^\top, \pmb{\eta}^\top)^\top \) which stacks the parameters in the outcome model and calibration equations. To simplify the notation, we use $\pmb{u}_i$, $\pmb{u}_i^*$, and $F_i$ to respectively denote $\pmb{u}_i(\pmb{\beta})$, $\pmb{u}_i^*(\pmb{\beta}^*)$, and $F_i(\pmb{\eta}) = \pmb{u}_i^{*\top}\pmb{\eta} + 1$ in this section.\\[-35pt]

\paragraph{(A) Pooled-sample calibration.}
When  microdata from both $s_1$ and $s_2$ are available, 
the stacked estimating equation system is defined as
\begin{equation}
\Psi(\pmb{\theta}) =\left(\begin{array}{c}
\pmb{U}(\pmb{\beta})\\[3pt]
\pmb{Q}(\pmb{\eta})
\end{array}\right)=
\left(\begin{array}{l}
\sum_{i\in \mathcal{U}}\delta_i^{(s_1)} w_i^{(s_1)} F_i\,\pmb{u}_i \\
\sum_{i\in \mathcal{U}}\delta_i^{(s_1)} w_i^{(s_1)} F_i\,\pmb{u}^*_i
- \sum_{i\in \mathcal{U}}\sum_{k=1}^2 f_k \delta_i^{(s_k)} w_i^{(s_k)} \pmb{u}^*_i
\end{array}
\right)
\label{combined_stack}
\end{equation}
where $f_k = n_k/n$ is the proportion of the sample $s_k$ in the pooled sample $S$, $k=1,2$.
At the FP parameters \(\pmb{\theta}_0=(\pmb{\beta}_{FP}^\top,\pmb{\eta}_{FP}^\top)^\top\),
the stacked estimating functions satisfy $E\{\Psi(\pmb{\theta}_0)\}=\pmb{0},$
with $E\{\pmb{U}(\pmb{\beta}_{FP})\}=\pmb{0},E\{\pmb{Q}(\pmb{\eta}_{FP})\}=\pmb{0}.$
Using a first-order Taylor expansion of \(\Psi(\pmb{\theta})\)
around \(\pmb{\theta}_0\), we obtain
\[
\widehat{\pmb{\theta}}-\pmb{\theta}_0
=
-\pmb{\psi}^{-1}(\pmb{\theta}_0)\Psi(\pmb{\theta}_0)
+o_p(n_1^{-1/2}),
\]
where
\[
\pmb{\psi}^{-1}(\pmb{\theta}) =
\begin{pmatrix}
\pmb{U}_{\pmb{\beta}}^{-1} & \pmb{b}\\
\pmb{0} & \pmb{Q}_{\pmb{\eta}}^{-1}
\end{pmatrix},
\]
with
\[
\pmb{U}_{\pmb{\beta}}
=E\left\{
\frac{\partial \pmb{U}(\pmb{\beta})}{\partial \pmb{\beta}^\top}\right\}
,\
\pmb{Q}_{\pmb{\eta}}
=
E\left\{\frac{\partial \pmb{Q}(\pmb{\eta})}{\partial \pmb{\eta}^\top}\right\}
,
\
\pmb{U}_{\pmb{\eta}}
=E\left\{
\frac{\partial \pmb{U}(\pmb{\eta})}{\partial \pmb{\eta}^\top}\right\},
\
\pmb{b}
=
-\pmb{U}_{\pmb{\beta}}^{-1}
\pmb{U}_{\pmb{\eta}}
\pmb{Q}_{\pmb{\eta}}^{-1}.
\] (details in Supplementary Materials Section 2.3). 
Then we have
\[
\mathrm{Var}(\widehat{\pmb{\theta}}) =
\psi^{-1}(\pmb{\theta}_0)\mathrm{Var}\{\Psi(\pmb{\theta}_0)\}\left\{\psi^{-1}(\pmb{\theta}_0)\right\}^{\!\top}+o(n_1^{-1}),
\]
where $\mathrm{Var}(\cdot)$ is with respect to the randomness of $\delta_i^{(s_1)}$ and $\delta_i^{(s_2)}$. 
Under Assumption (A.2), $\mathrm{Var}\{\Psi(\pmb{\theta})\}$ can be decomposed as follows
\begin{equation}
\begin{aligned}
&\mathrm{Var}\{\Psi(\pmb{\theta})\} =
\begin{pmatrix}
\pmb{V}_{11}^{(s_1)}& \pmb{V}_{12}^{(s_1)}\\
\pmb{V}_{12}^{(s_1)\top}& \pmb{V}_{22}^{(s_1)} + \pmb{V}_{22}^{(s_2)}
\end{pmatrix},
\label{combined_phi}
\end{aligned}
\end{equation}
where \\
$\pmb{V}_{11}^{(s_1)}=\mathrm{Var}_{s_1}\left\{\sum_{i\in \mathcal{U}}\delta_i^{(s_1)} w_i^{(s_1)} F_i\pmb{u}_i\right\}$,\\
$\pmb{V}_{22}^{(s_1)}=\mathrm{Var}_{s_1}\left\{\sum_{i\in \mathcal{U}}\delta_i^{(s_1)} w_i^{(s_1)} (F_i-f_1)\pmb{u}_i^*\right\},$\\  
$\pmb{V}_{22}^{(s_2)}=f_2^2\mathrm{Var}_{s_2}\left\{\sum_{i\in \mathcal{U}}\delta_i^{(s_2)} w_i^{(s_2)}\pmb{u}_i^*\right\}$,\\
$\pmb{V}_{12}^{(s_1)}=\mathrm{Cov}_{s_1}\left\{\sum_{i\in \mathcal{U}}\delta_i^{(s_1)} w_i^{(s_1)} F_i\pmb{u}_i,\ \sum_{i\in \mathcal{U}}\delta_i^{(s_1)} w_i^{(s_1)} (F_i-f_1)\pmb{u}_i^*\right\},$\\
with $\mathrm{Var}_{s_k}$ and $\mathrm{Cov}_{s_k}$ respectively denoting the variance and covariance due to the sampling indicators $\delta^{(s_k)},\ k=1,2$, accounting for the general stratified multistage clustering designs with unequal inclusion probabilities. 
Accordingly, the FP variance of $\widehat{\pmb{\beta}}_\text{p.clb}$ can be approximately obtained by 
\begin{equation}
\begin{aligned}
\mathrm{Var}(\widehat{\pmb{\beta}}_{\mathrm{p.clb}})
=
&\ \pmb{U}_{\pmb{\beta}}^{-1}
\pmb{V}_{11}^{(s_1)}
\pmb{U}_{\pmb{\beta}}^{-1\top}  \\
&+ \pmb{b}\,
\pmb{V}_{12}^{(s_1)\top}
\pmb{U}_{\pmb{\beta}}^{-1\top}
+ \pmb{U}_{\pmb{\beta}}^{-1}
\pmb{V}_{12}^{(s_1)}
\pmb{b}^{\top} \\
&+ \pmb{b}\,
\pmb{V}_{22}^{(s_1)}
\pmb{b}^{\top}
+ \pmb{b}\,
\pmb{V}_{22}^{(s_2)}
\pmb{b}^{\top}
+ o(n_1^{-1}).
\end{aligned}
\label{var_calib}
\end{equation}
The plug-in sample variance estimator of $\mathrm{Var}\{\widehat{\pmb{\beta}}_\text{p.clb}\}$ can be obtained by replacing the unknown FP quantities in $\mathrm{Var}(\widehat{\pmb{\beta}}_{\mathrm{p.clb}})$ with their sample-weighted estimates (Supplementary Materials Section~2.5).\\[-30pt]

\paragraph{(B) External-sample calibration.}
The FP variance of $\widehat{\pmb{\beta}}_\text{e.clb}$, denoted by $\mathrm{Var}(\widehat{\pmb{\beta}}_{\mathrm{e.clb}})$, can be derived similarly to $\mathrm{Var}(\widehat{\pmb{\beta}}_{\mathrm{p.clb}})$ but uses a different calibration equation
$
\pmb{Q}^*(\pmb{\eta})
=
\sum_{i\in \mathcal{U}}\delta_i^{(s_1)} w_i^{(s_1)} F_i \pmb{u}_i^*
- \sum_{i\in \mathcal{U}}\delta_i^{(s_2)} w_i^{(s_2)} \pmb{u}_i^*
$
to replace $\pmb{Q}(\pmb{\eta})$ in \eqref{combined_phi} to account for the randomness due to sampling $s_2$. Accordingly, $\mathrm{Var}(\widehat{\pmb{\beta}}_{\mathrm{e.clb}})$ can be approximately obtained by
\begin{equation}
\label{var_ecalib}
\begin{aligned}
\mathrm{Var}(\widehat{\pmb{\beta}}_{\mathrm{e.clb}})
\approx&\
\pmb{U}_{\pmb{\beta}}^{-1}\left\{\pmb{V}_{11}^{(s_1)}+\pmb{U}_{\pmb{\eta}}\mathrm{Var}_{s_2}(\widehat{\pmb{\beta}}^*)\pmb{U}_{\pmb{\eta}}^\top\right\}\pmb{U}_{\pmb{\beta}}^{-1\top}
+ \pmb{b}^*\,\pmb{V}_{12}^{*(s_1)\top}\pmb{U}_{\pmb{\beta}}^{-1\top}
\\
&+ \pmb{U}_{\pmb{\beta}}^{-1}\pmb{V}_{12}^{*(s_1)}\pmb{b}^{*\top}+ \pmb{b}^*\,\pmb{V}_{22}^{*(s_1)}\pmb{b}^{*\top},
\end{aligned}
\end{equation}
where $\pmb{b}^*= -\,\pmb{U}_{\pmb{\beta}}^{-1}\pmb{U}_{\pmb{\eta}} \pmb{Q}_{\pmb{\eta}}^{*-1}$, $\pmb{V}_{22}^{*(s_1)}=\mathrm{Var}_{s_1}\left\{\sum_{i\in \mathcal{U}}\delta_i^{(s_1)} w_i^{(s_1)} F_i\pmb{u}_i^*\right\}$, and
$\pmb{V}_{12}^{*(s_1)}$ $=\mathrm{Cov}_{s_1}\left\{\sum_{i\in \mathcal{U}}\delta_i^{(s_1)} w_i^{(s_1)} F_i\pmb{u}_i,\ \sum_{i\in \mathcal{U}}\delta_i^{(s_1)} w_i^{(s_1)} F_i\pmb{u}_i^*\right\}$. Notice that $\mathrm{Var}(\widehat{\pmb{\beta}}_{\mathrm{e.clb}})$ above only requires $\mathrm{Var}_{s_2}(\widehat{\pmb{\beta}}^*)$ but not microdata from $s_2$. The detailed derivation and the sample plug-in variance estimator are provided in Supplementary Materials Section~2.6.
\section{Simulation Study}

We evaluate the finite-sample performance of the proposed pooled- and external-sample calibration estimators in terms of their approximate unbiasedness of $\pmb{\beta}_{FP}$ and efficiency under various scenarios with different sample sizes of $s_2$ and the correlation between the expensive predictor and its predictions.

\subsection{Data Generation}
We generated a FP of size $N = 1{,}000{,}000$, with values for three covariates $X_1$, $X_2$, and $X_3$ generated as follows: 
$X_1 \sim \text{Bernoulli}(0.3)$, and 
$(X_2, X_3)^\top \sim \mathcal{N}(\pmb{0}, \pmb{\Sigma})$, where $\pmb{\Sigma} = \begin{bmatrix} 1 & 0.1 \\ 0.1 & 1 \end{bmatrix}$. 
We consider $X_1$ and $X_2$ as inexpensive covariates and $X_3$ as the expensive covariate. 
In addition, we generated two surrogates of $X_3$, denoted by $X_3^{*(1)}$ and $X_3^{*(2)}$, from 
$ X_{3}^{*(j)} = X_3 + a_j \cdot \log(|X_3 + \varepsilon|)$, $j = 1, 2$, 
where $\varepsilon \sim \mathcal{N}(0, 0.5)$ and a constant $a_j$ that controls the correlation between $X_3$ and $X_3^{*(j)}$. These two variables play the role of the ancillary variables \(\pmb Z\) in the proposed calibration framework.
We set $a_1 = 1.55$ and $a_2 = 0.67$, leading to empirical correlations between $X_3$ and the two surrogates of 
$\rho_{X_3, X_{3}^{*(1)}} \approx 0.50$ and $\rho_{X_3, X_{3}^{*(2)}} \approx 0.80$, respectively.

To mimic the generation of a survey sampling frame in the FP to real data,  
we partition the FP into $M = 10{,}000$ clusters, each containing approximately $N/M$ individuals depending on their values of $\pmb{X}$. 
We then generate values of the outcome variable $Y$ by $Y_i \sim \text{Bernoulli}(p_i)$ for $i \in \mathcal{U}$, with the intra-cluster correlation (ICC) for $Y$ set to 0.1 \citep{kerry1998intracluster,thompson2012icc}, where 
$p_i = \text{expit}\{\beta_0 + \beta_1 X_{1,i} + \beta_2 X_{2,i} + \beta_3 X_{3,i} + \beta_{1,3} (X_{1,i}\cdot X_{3,i})\}$, 
and the true value of the parameter vector is $\pmb{\beta} = (\beta_{0}, \beta_{1}, \beta_2, \beta_{3}, \beta_{1,3})^\top = (-3.5, 1.0, 0.5, 0.5, 0.2)^\top$.

We independently draw two probability samples from the FP $B = 1000$ times: an internal sample ($s_1$) serving as the primary internal analytic dataset, and an external sample ($s_2$). 
Both samples are randomly selected using two-stage cluster sampling designs with different selection probabilities at each stage. 
The details of the sampling designs and the generation of the sample inclusion probabilities $\{\pi_i^{(1)}, \pi_i^{(2)}, i \in \mathcal{U}\}$ are described in Supplementary Materials Section~3. 

The internal sample $s_1$ consists of $m_{1} = 60$ clusters and a total sample size of $n_{1} = 2{,}000$. 
The number of clusters in the external sample $s_2$ varies across $m_{2} \in \{60,\ 180,\ 300,\ 600\}$, yielding total sample sizes $n_{2} \in \{2{,}000,\ 6{,}000,\ 10{,}000,\ 20{,}000\}$, respectively.

\subsection{Results}

Table~\ref{tab:bias_vr_full} shows the results of the calibration estimators using different predictions of $X_3$ in the surrogate model when $n_2 = 6000$. The results using the reduced model excluding $X_3$ and its interaction with $X_1$, as well as the full outcome model, are also reported as the worst- and best-case scenarios for comparison. 

Both pooled- and external-sample calibration approaches yield approximately unbiased estimators of $\pmb{\beta}$ and achieve substantial efficiency gains over the estimator using only $s_1$. The efficiency gain in estimating all components of $\pmb{\beta}$ increases when the correlation between $X_3$ and $X_{3}^{*(j)}$ is greater. However, the magnitude of efficiency improvement varies across the components of $\pmb{\beta}$. Calibration for $\beta_{x_2}$ consistently achieves the largest efficiency gain, reducing the variance of the $s_1$ estimator by 73\%--76\% and by 65\%--68\% with and without access to microdata from $s_2$, respectively. Calibration estimators of $\beta_{x_1}$ are also highly efficient, but their gains are slightly weaker
\FloatBarrier
\begin{table}[H]
\caption{Simulation results for the calibration estimators of regression coefficients using different surrogate prediction models ($n_2=6000$).}
\label{tab:bias_vr_full}

\small
\setlength{\tabcolsep}{3.5pt}
\renewcommand{\arraystretch}{0.9}

\begin{tabular}{lccccccccc}
\toprule
 & \multicolumn{4}{c}{\textbf{Pooled Calibration}} && \multicolumn{4}{c}{\textbf{External Calibration}} \\
\cmidrule(lr){2-5} \cmidrule(lr){7-10}
 & Bias$^a$ (\%) & VR$^b$ & CR$^c$ (\%) & RelVar$^d$ && Bias (\%) & VR & CR (\%) & RelVar \\
\midrule

\multicolumn{10}{l}{\textbf{Surrogate model}: $Y \sim X_1,  X_2$ (Reduced prediction model)}  \\
$\beta_{x_1}$     & -0.54 & 1.01 & 96.3 & 0.40 && -0.36 & 1.03 & 94.9 & 0.45 \\
$\beta_{x_2}$     &  0.30 & 1.06 & 96.0 & 0.27 && -0.73 & 1.05 & 95.2 & 0.35 \\
$\beta_{x_3}$     & -0.31 & 1.04 & 95.1 & 1.01 && -0.30 & 1.04 & 95.2 & 1.01 \\
$\beta_{x_1:x_3}$ &  4.80 & 0.97 & 93.8 & 1.00 &&  4.89 & 0.97 & 93.9 & 1.01 \\
\midrule
\multicolumn{10}{l}{\textbf{Surrogate model}: $Y \sim X_1,  X_2,X_3^{*(1)}, X_1\cdot X_3^{*(1)}$} \\
$\beta_{x_1}$     & -0.41 & 1.01 & 95.1 & 0.31 && -0.19 & 1.02 & 95.0 & 0.37 \\
$\beta_{x_2}$     &  0.23 & 1.07 & 95.8 & 0.26 &&  0.69 & 1.07 & 95.4 & 0.33 \\
$\beta_{x_3}$     & -1.95 & 1.01 & 95.2 & 0.69 && -2.42 & 1.01 & 94.8 & 0.75 \\
$\beta_{x_1:x_3}$ &  3.44 & 0.95 & 93.9 & 0.67 &&  3.51 & 0.94 & 94.1 & 0.73 \\
\midrule
\multicolumn{10}{l}{\textbf{Surrogate model}: $Y \sim X_1,  X_2,X_3^{*(2)}, X_1\cdot X_3^{*(2)}$} \\
$\beta_{x_1}$     & -0.29 & 1.02 & 94.3 & 0.27 && -0.03 & 1.02 & 94.7 & 0.34 \\
$\beta_{x_2}$     &  0.11 & 1.08 & 95.7 & 0.25 &&  0.54 & 1.07 & 95.6 & 0.33 \\
$\beta_{x_3}$     & -0.58 & 1.03 & 95.3 & 0.45 && -0.54 & 1.03 & 94.9 & 0.53 \\
$\beta_{x_1:x_3}$ &  1.54 & 0.96 & 94.8 & 0.45 &&  1.13 & 0.95 & 94.4 & 0.52 \\
\midrule
\multicolumn{10}{l}{\textbf{Surrogate model}: $Y \sim X_1,  X_2,X_3, X_1\cdot X_3$ (Full prediction model)} \\
$\beta_{x_1}$     & -0.03 & 1.04 & 95.9 & 0.22 &&  0.33 & 1.00 & 94.9 & 0.31 \\
$\beta_{x_2}$     &  0.10 & 1.09 & 95.6 & 0.24 &&  0.52 & 1.08 & 95.8 & 0.32 \\
$\beta_{x_3}$     &  0.51 & 1.04 & 95.0 & 0.26 &&  0.96 & 1.01 & 95.0 & 0.36 \\
$\beta_{x_1:x_3}$ & -0.86 & 0.99 & 94.7 & 0.25 && -1.99 & 0.98 & 94.9 & 0.33 \\
\bottomrule
\end{tabular}

\vskip 2mm
\renewcommand{\arraystretch}{0.5}
{\footnotesize
$^{a}$~Relative bias: the difference between the simulation average of the point estimates and the true FP value divided by the the true FP value, times $100\%$.

\noindent
$^{b}$~Variance ratio (VR): the ratio of the average TL variance estimator to the empirical simulation variance.

\noindent
$^{c}$~95\% coverage rate (CR): percentage of the 95\% confidence intervals using the TL standard error that cover the true parameter.

\noindent
$^{d}$~Relative variance (RelVar): the ratio of the empirical variance of the calibrated estimator to that of the $s_1$ estimator.
}
\end{table}

\noindent than those of $\beta_{x_2}$. This is because $X_1$ is involved in an interaction term with the expensive covariate $X_3$ in the outcome model of interest, so the efficiency gain for estimating $\beta_{x_1}$ depends on how strongly the surrogate variable correlates with $X_3$.

For $\beta_{x_3}$ and $\beta_{x_1:x_3}$, the efficiency gains of the calibration estimators depend critically on the correlation between $X_3$ and its surrogate. Under the reduced surrogate model with no prediction of $X_3$, no efficiency improvement is observed, whereas including the prediction of $X_3$ leads to substantial gains, especially when $X_3^{*(2)}$ is used due to its strong correlation with $X_3$.




\noindent A similar pattern of efficiency gain holds for the estimator of the interaction effect $\beta_{x_1:x_3}$.

It is also shown in Table~\ref{tab:bias_vr_full} that the proposed TL variance estimator performs well, with the variance ratios all close to 1 and the coverage rates of the confidence intervals close to the nominal level. 

Table~\ref{tab:simulation_theta} shows that the pooled- and external-sample calibration estimators are asymptotically equivalent in efficiency to the augmented estimators $\widehat{\pmb{\beta}}_\text{p}(\pmb{\Theta})$ and $\widehat{\pmb{\beta}}_\text{e}(\pmb{\Theta})$, respectively, which use the unknown population influence functions to calculate $\pmb{\Theta}$. Notice that the calibration estimators perform substantially differently from the augmented estimators $\widehat{\pmb{\beta}}_\text{p}(\widehat{\pmb{\Theta}})$ and $\widehat{\pmb{\beta}}_\text{e}(\widehat{\pmb{\Theta}})$, which use the sample  plug-in influence functions in   $\widehat{\pmb{\Theta}}$ to substitute the unknown $\pmb{\Theta}$ in $\widehat{\pmb{\beta}}_\text{p}(\pmb{\Theta})$. Using the sample plug-in $\widehat{\pmb{\Theta}}$ not only lead to lower efficiency gain but also induce extra bias especially when the the expensive covariate is highly correlated with its prediction in the external-calibration setting. These empirical results are aligned with the theoretical justifications in Supplementary Materials Section~1.10.

\begin{table}[H]
\caption{Comparison of calibration and plug-in estimators ($n_2=6000$).}
\label{tab:simulation_theta}

\small
\setlength{\tabcolsep}{5pt}
\renewcommand{\arraystretch}{0.9}

\begin{tabular}{lccccccccccc}
\toprule
& \multicolumn{5}{c}{\textbf{Pooled Calibration}}&
& \multicolumn{5}{c}{\textbf{External Calibration}} \\
\cmidrule(lr){2-6}\cmidrule(lr){8-12}

& \multicolumn{2}{c}{Bias$^a$ (\%)}&
& \multicolumn{2}{c}{RV$^b$}&
& \multicolumn{2}{c}{Bias$^a$ (\%)}&
& \multicolumn{2}{c}{RV$^b$} \\
\cmidrule(lr){2-3}\cmidrule(lr){5-6}\cmidrule(lr){8-9}\cmidrule(lr){11-12}

& $\widehat{\pmb\beta}_\text{p}(\pmb\Theta)$ & $\widehat{\pmb\beta}_\text{p}(\widehat{\pmb\Theta})$ && $\widehat{\pmb\beta}_\text{p}(\pmb\Theta)$ & $\widehat{\pmb\beta}_\text{p}(\widehat{\pmb\Theta})$&
& $\widehat{\pmb\beta}_\text{e}(\pmb\Theta)$ & $\widehat{\pmb\beta}_\text{e}(\widehat{\pmb\Theta})$ && $\widehat{\pmb\beta}_\text{e}(\pmb\Theta)$ & $\widehat{\pmb\beta}_\text{e}(\widehat{\pmb\Theta})$ \\

\midrule

\multicolumn{12}{l}{\textbf{Surrogate model}: $Y \sim X_1,  X_2$ (Reduced prediction model)}\\
$\beta_{x_1}$     & -0.44 & -0.96 && 1.02 & 1.02 && -0.30 & -1.18 && 1.02 & 1.16 \\
$\beta_{x_2}$     &  -0.13 & -0.79 && 1.00 & 1.03 &&  0.34 & -0.82 && 1.00 & 1.13 \\
$\beta_{x_3}$     & -0.24 & -0.28 && 1.04 & 1.00 && -0.23 & -0.28 && 1.06 & 1.00 \\
$\beta_{x_1:x_3}$ &  2.37 &  2.58 && 1.03 & 1.00 &&  2.25 &  2.53 && 1.04 & 1.00 \\

\midrule
\multicolumn{12}{l}{\textbf{Surrogate model}: $Y \sim X_1,  X_2,X_3^{*(1)}, X_1\cdot X_3^{*(1)}$}  \\
$\beta_{x_1}$     & -0.31 & -0.88 && 1.06 & 1.19 && -0.04 & -1.19 && 1.08 & 1.61 \\
$\beta_{x_2}$     &  -0.16 & -0.97 && 1.00 & 1.03 && 0.36 & -1.06 && 1.00 & 1.16 \\
$\beta_{x_3}$    & -2.06 & -2.62 && 1.06 & 1.07 &&  -2.50 & -3.55 && 1.05 & 1.17 \\
$\beta_{x_1:x_3}$ &  0.94 &  -0.40 && 1.04 & 1.05 &&   0.49 &  -2.13 && 1.05 & 1.14 \\

\midrule
\multicolumn{12}{l}{\textbf{Surrogate model}: $Y \sim X_1,  X_2,X_3^{*(2)}, X_1\cdot X_3^{*(2)}$} \\
$\beta_{x_1}$     & -0.14 & -0.72 && 1.02 & 1.21 &&  0.16 & -0.92 && 1.04 & 1.63 \\
$\beta_{x_2}$     &  -0.28 & -1.22 && 1.00 & 1.04 &&  0.23 & -1.40 && 1.00 & 1.16 \\
$\beta_{x_3}$     & -0.17 & -0.90 && 1.02 & 1.11 && -0.19 & -1.05 && 1.02 & 1.25 \\
$\beta_{x_1:x_3}$ &  -1.33 & -4.04 && 1.03 & 1.07 &&  2.09 & -7.33 && 1.04 & 1.20 \\

\midrule
\multicolumn{12}{l}{\textbf{Surrogate model}: $Y \sim X_1,  X_2,X_3, X_1\cdot X_3$ (Full prediction model)} \\
$\beta_{x_1}$     &  0.05 & -0.53 && 1.00 & 1.24 &&  0.37 & -0.63 && 1.00 & 1.70 \\
$\beta_{x_2}$     & -0.27 &  -1.30 && 1.00 & 1.03 &&  0.23 & -1.56 && 1.00 & 1.15 \\
$\beta_{x_3}$     & 0.18 &  0.74 && 1.00 & 1.26 &&  0.51 &  -1.09 && 1.00 & 1.53 \\
$\beta_{x_1:x_3}$ & -2.65 & -6.37 && 1.00 & 1.15 && -3.71 & -10.27 && 1.00 & 1.36 \\
\bottomrule
\end{tabular}
\vskip 2mm
\renewcommand{\arraystretch}{0.9}
{\footnotesize 
$^{a}$~Relative bias: Relative bias: the difference between the simulation average of the point estimates and the true FP value divided by the the true FP value, times $100\%$.

\noindent
$^{b}$~Relative variance (RV): the empirical variance of augmented estimators shown in the column divided by that of the corresponding calibration estimator, $\widehat{\pmb{\beta}}_{\mathrm{p.clb}}$ or $\widehat{\pmb{\beta}}_{\mathrm{e.clb}}$.
}
\end{table}
\newpage
\begin{figure}[ht]
\centering
\begin{tikzpicture}
\node[anchor=south west] (img) at (0,0)
  {\includegraphics[width=0.9\textwidth]{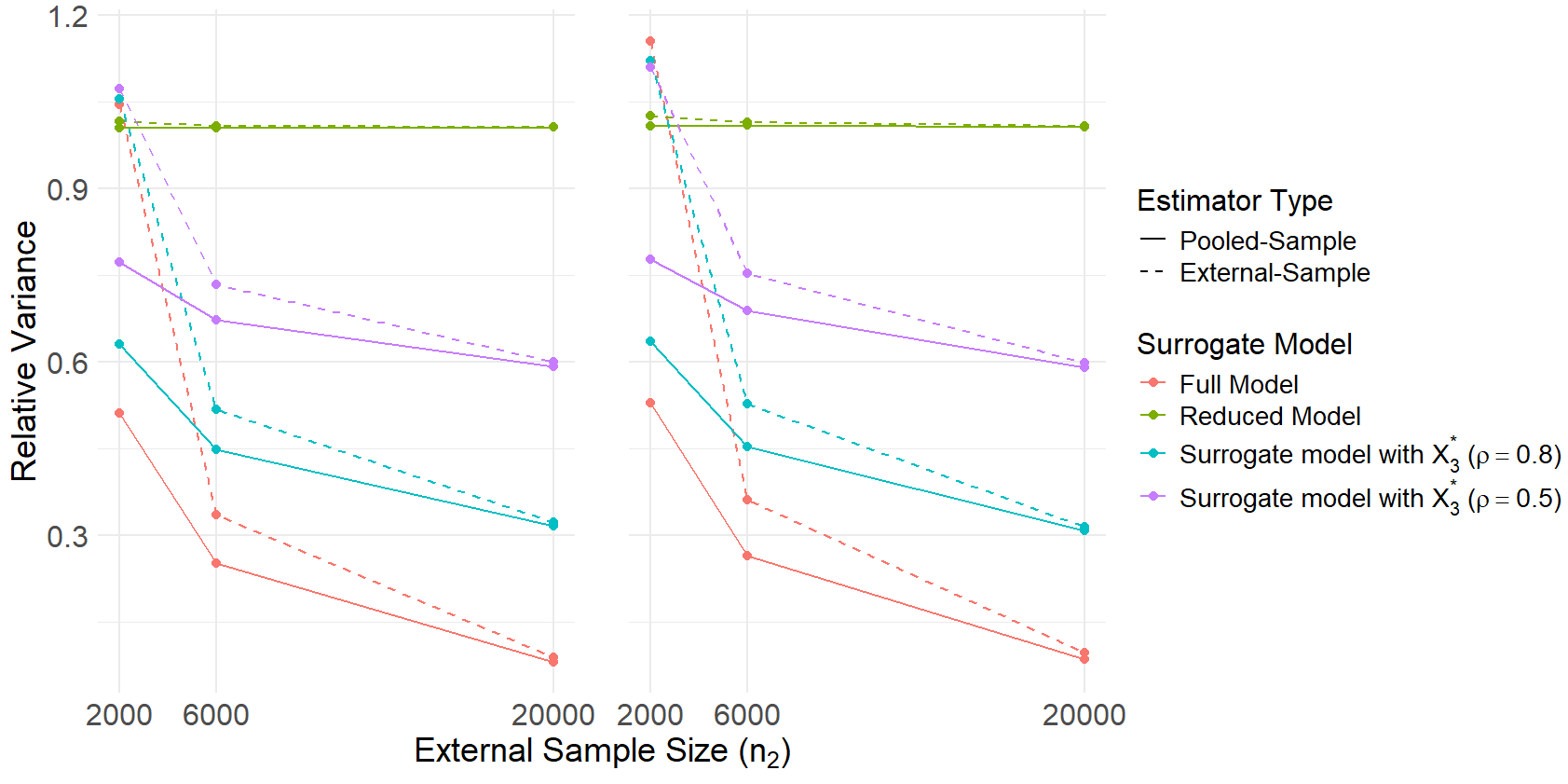}};
\begin{scope}[x={(img.south east)},y={(img.north west)}]
  \node at (0.25,1.02) {\textbf{(a)}};
  \node at (0.55,1.02) {\textbf{(b)}};
\end{scope}
\end{tikzpicture}

\caption{Relative variance of the calibration estimator for (a) $\beta_{x_3}$ and (b) $\beta_{x_1:x_3}$ compared to the $s_1$ estimator with increased external sample size.}
\label{fig:3}
\end{figure}

\noindent Figure~\ref{fig:3} illustrates that both pooled- and external-sample calibration estimators of $\beta_{x_3}$ and $\beta_{x_1:x_3}$ achieve increased efficiency gains as $n_2$ grows. The pooled-sample calibration estimator is consistently more efficient than the external-sample calibration estimator under the same surrogate model due to the larger pooled sample size $n > n_2$. The additional efficiency gain of the pooled-sample calibration is more substantial when $n_2$ is small, but diminishes when $n_2$ is much larger than $n_1$.

We note that when $n_2$ is small, the external-sample calibration estimator can be slightly less efficient than the internal-sample estimator, as shown by the relative variance exceeding one in Figure~\ref{fig:3}(b). This occurs because the external-sample calibration estimator relies on the summary estimator $\widehat{\pmb{\beta}}^*_{s_2}$, whose sampling variability is carried into the calibration equations. When the external sample is not sufficiently large or the surrogate information for the expensive covariate is weak, this additional variability may outweigh the variance reduction from calibration. As $n_2$ increases, the variability of the external summary estimator decreases, and the external-sample calibration estimator becomes more efficient than the internal-sample estimator.
\section{Real Data Application: NHANES and NHIS}



We apply the proposed methods to estimate the association between the prospective 10-year all-cause mortality and total body fat, adjusted for demographic characteristics and lifestyles. The data are from the National Health and Nutrition Examination Survey (NHANES) and the National Health Interview Survey (NHIS), conducted by the U.S. National Center for Health Statistics. Both NHANES and NHIS are repeated cross-sectional surveys that apply stratified multistage cluster sampling designs with differential selection probabilities at each stage to randomly select samples that represent the civilian, non-institutionalized U.S. population. They are linked to the National Death Index for mortality follow-up through 2019. 

The expensive covariate of interest, total body fat, was measured via dual-energy X-ray absorptiometry (DXA) in the NHANES physical examinations in four biennial survey cycles between 1999 and 2006. This measurement was only collected among participants whose ages at the time of the survey were between 18 and 69 years, with body weight below 300 pounds, height below 6.2 feet, and who were not pregnant. We found that the distribution of total body fat by 10-year mortality status in the 1999--2000 survey cycle differs substantially from that in later cycles (Figure S1 in Supplementary Materials), which may violate Assumption (A.1). Therefore, we exclude this cycle in the analysis and focus on the three cycles between 2001 and 2006.




We use sample collected by NHANES 2003--2004 ($n_1 = 1{,}543$) as the internal sample ($s_1$) to fit the risk model for 10-year all-cause mortality. In addition to total body fat, we also include age (in years), sex, race/ethnicity, smoking status, intensity of drinking (measured by the frequency and amount of alcohol consumption), and physical activity (measured by Metabolic Equivalents) as covariates.

The external sample is from contemporaneous NHIS 2003--2004, restricted to individuals who are eligible for DXA ($s_2$ with $n_2 = 24{,}086$). The distribution of the covariates in the weighted NHIS are very close to those in weighted NHANES (Table S1 in Supplementary Material). Because DXA is not used in NHIS, we use self-reported Body Mass Index (BMI) as a prediction for total body fat. Notice that the calibration methods require no measurement bias or data harmonization issues between the internal and the external samples. In our data example, NHANES participants were aware that their height and weight would be measured during the physical examination and therefore tended to report BMI values that were closer to their measured BMI than did NHIS participants. (Figure S2 in Supplementary Materials). The slight different distributions of self-reported BMI may induce potential bias. Therefore, for comparison purposes, we also consider another \lq\lq external data source\rq\rq\ that combines data from two NHANES cycles, 2001--2002 and 2005--2006, denoted by $s_3$ with $n_3 = 3{,}107$. As $s_1$ and $s_3$ can be treated as independently sampled by the same study within a similar time period, integrating data from $s_1$ and $s_3$ satisfies assumptions (A.1), (A.2) with less concerns about different measurement errors or data harmonization issues. We mask the observed total body fat in $s_3$ and only use self-reported BMI as a prediction for total body fat. 

To evaluate the performance of the pooled- and external-sample calibration approaches, we use a \lq\lq benchmark\rq\rq\  estimators obtained from the combined sample of three NHANES 2001--2006 cycles, where the total body fat is observed for all individuals.

\begin{figure}[H]
\centering
\includegraphics[width=\textwidth]{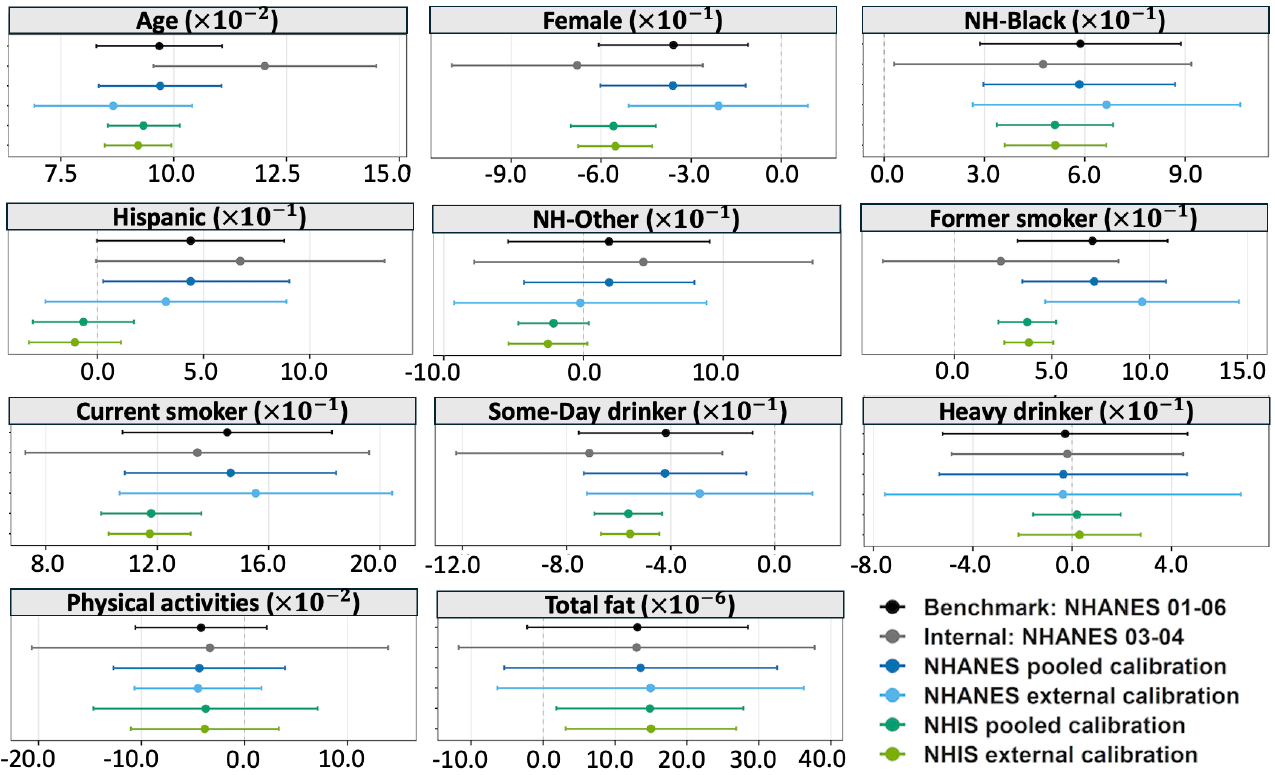}
\caption{Point estimates and 95\% confidence intervals of log-odds ratios for 10-year all-cause mortality.}
\label{fig:realdata-ci}
\end{figure}

\noindent As shown in Figure~\ref{fig:realdata-ci}
(detailed numbers shown in Table S2 in Supplementary Materials), compared to the estimates obtained from one cycle of NHANES, both calibration approaches generally yield narrower confidence intervals than the internal-sample analysis. For example, for the log-OR of the total body fat, the confidence interval width decreases from 49.39 in the internal-sample analysis to 25.96 and 23.67 when using NHIS as the external sample under pooled- and external-sample calibration, respectively.

When using the combined 2001--2002 and 2005--2006 cycles of NHANES as the external data, the calibration estimates are close to the benchmark, reflecting consistency between the internal and external samples. The confidence intervals for the external-sample calibration estimators are generally slightly wider than those of the pooled-sample calibration estimators, suggesting a small efficiency loss resulting from the use of summary statistics rather than external-sample microdata when the external and internal sample sizes are similar.

When using the contemporaneous NHIS as the external data source, the calibration estimates slightly deviate from the benchmark, possibly due to measurement discrepancies and data harmonization issues between the two surveys. For example, differences in self-reported health-related variables across NHANES and NHIS may introduce small discrepancies, as the two surveys differ in data collection settings and measurement protocol. \citep{nchs2013nhanes,nchs2004nhis}. Despite the small differences in the point estimates, the large NHIS sample size leads to substantial additional variance reduction compared to the results using other cycles of NHANES as the external sample. Also, due to the large NHIS sample size, the pooled- and external-sample calibration estimators achieve similar efficiency. Both methods show a significant positive association between total body fat and all-cause mortality after adjusting for other covariates, which is not captured by the benchmark. In addition, all calibration methods, using either NHANES or NHIS as external data, show a significant effect of former smoking on increasing the risk of all-cause mortality, which is not observed in the one-cycle 2003--2004 NHANES internal sample, mainly due to the small sample size of former smokers.

\section{Discussion}

We develop a model-assisted calibration framework for regression analysis when key variables are observed only in a small internal probability sample and auxiliary information is available from a much larger external survey. By using score functions as calibration auxiliary variables, the proposed method achieves efficiency gains in estimating regression parameters while preserving design consistency under complex sampling designs. The calibration method can be applied with or without access to the microdata of the larger external sample. We show that the calibration estimators can achieve similar efficiency to the augmented estimators, which require unknown population quantities. We also provide variance estimators for design-based inference that appropriately account for variability due to the randomness of both internal and external survey samples.


Simulation results demonstrate that the efficiency gains of the calibration estimators for all association parameters in the outcome model depend on the sample size of the external data. The efficiency of the estimated effects for the expensive covariate and the interaction terms further depends on the extent of the correlation between the expensive covariate and its prediction. External-sample calibration can be less efficient than pooled-sample calibration due to relying only on summary statistics, but this efficiency loss vanishes when the external sample size is much larger than the internal sample size.


In the data application of risk estimation for all-cause mortality using data from NHANES and NHIS, the proposed calibration method substantially improved the efficiency in estimating the log odds ratio of total body fat, which is only measured in NHANES, and detected significant effects of former smoking and total body fat that were not captured by the NHANES-only analysis. This application also illustrates the robustness of the proposed framework under mild departures from the ideal assumptions. Specifically, using other NHANES cycles as external data may slightly violate Assumption~(A.1) because survey cycles from different years may represent slightly different FPs, whereas using NHIS as the external data source may be subject to minor data harmonization issues arising from differences in the measurement of some variables across the two surveys. Nevertheless, both external data sources led to substantial efficiency gains with only modest changes in point estimates, suggesting that the proposed calibration approach can remain useful in practical settings where the assumptions are approximately, but not perfectly, satisfied.

Note that although the calibration methods are proposed for survey data integration, they can also be applied to integrating data without complex sampling designs, such as data from medical studies, epidemiological cohort studies, or electronic health records, provided that they represent the same population. The methods can also be naturally extended to settings with multiple external data sources.
The proposed methods have several limitations. First, the efficiency gain achieved by the calibration method can be relatively small when the correlation between the expensive covariate and its predictor is low. Future research is needed to improve the working model for predicting the expensive covariate using statistical learning techniques. 
Second, the methods require that the internal and external samples represent the same target FP. This strong assumption may restrict the applicability of the methods to integrating probability samples from different populations, such as survey data collected from different countries, or integrating probability samples with volunteer-based samples. Propensity weighting approaches can be considered to handle heterogeneous data distributions in the internal and external samples when microdata from both samples are available \citep{wang2025using}. Accounting for heterogeneous data distributions when only summary-level external data are available requires further research. 
Third, the calibration methods treat the external sample as a relatively reliable data source with less variability than the internal sample, which may result in limited efficiency gain or even efficiency loss when the external sample size is much smaller than that of the internal sample. Accounting for the randomness of the external sample when generating the calibration weights is an important direction for future research.

\nocite{*}
\bibliographystyle{apalike}   
\bibliography{references}     

\begin{thebibliography}{}

\bibitem[Angelopoulos et~al., 2023]{angelopoulos2023prediction}
Angelopoulos, A.~N., Bates, S., Fannjiang, C., Jordan, M.~I., and Zrnic, T. (2023).
\newblock Prediction-powered inference.
\newblock {\em Science}, 382(6671):669--674.

\bibitem[Breslow and Holubkov, 1997]{breslowholubkov1997}
Breslow, N.~E. and Holubkov, R. (1997).
\newblock Weighted likelihood for two-phase stratified samples.
\newblock {\em Biometrics}, 53(3):1134--1146.

\bibitem[Breslow et~al., 2009]{breslow2009}
Breslow, N.~E., Lumley, T., Ballantyne, C.~M., Chambless, L.~E., and Kulich, M. (2009).
\newblock Improved horvitz--thompson estimation of model parameters from two-phase stratified samples: Applications in epidemiology.
\newblock {\em Statistics in Biosciences}, 1:32--49.

\bibitem[Chatterjee et~al., 2016]{chatterjee2016constrained}
Chatterjee, N., Chen, Y.-H., Maas, P., and Carroll, R.~J. (2016).
\newblock Constrained maximum likelihood estimation for model calibration using summary-level information from external big data sources.
\newblock {\em Journal of the American Statistical Association}, 111(513):107--117.

\bibitem[Chen et~al., 2020]{chen2020}
Chen, Y., Li, P., and Wu, C. (2020).
\newblock Doubly robust inference with nonprobability survey samples.
\newblock {\em Journal of the American Statistical Association}, 115(532):2011--2021.

\bibitem[Deville and S{\"a}rndal, 1992]{deville1992calibration}
Deville, J.-C. and S{\"a}rndal, C.-E. (1992).
\newblock Calibration estimators in survey sampling.
\newblock {\em Journal of the American statistical Association}, 87(418):376--382.

\bibitem[Hu et~al., 2023]{hu2023}
Hu, T., Ning, Y., and Tchetgen~Tchetgen, E. (2023).
\newblock Semiparametric data fusion: Efficiency and paradoxes with noisy or biased summary sources.
\newblock {\em Journal of the Royal Statistical Society: Series B}, 85(5):1214--1241.

\bibitem[Kerry and Bland, 1998]{kerry1998intracluster}
Kerry, S.~M. and Bland, J.~M. (1998).
\newblock The intracluster correlation coefficient in cluster randomisation.
\newblock {\em BMJ}, 316(7142):1455.

\bibitem[Lumley et~al., 2011]{lumley2011}
Lumley, T., Shaw, P.~A., and Dai, J.~Y. (2011).
\newblock Connections between survey calibration, missing data, and semiparametric models.
\newblock {\em International Statistical Review}, 79(2):200--220.

\bibitem[{National Center for Health Statistics}, 2004]{nchs2004nhis}
{National Center for Health Statistics} (2004).
\newblock 2003 national health interview survey (nhis) public use data release: Survey description.
\newblock Technical report, National Center for Health Statistics.

\bibitem[{National Center for Health Statistics}, 2013]{nchs2013nhanes}
{National Center for Health Statistics} (2013).
\newblock National health and nutrition examination survey: Analytic guidelines, 1999--2010.
\newblock Technical Report Series 2, No. 161, National Center for Health Statistics.

\bibitem[Prentice, 1986]{prentice1986}
Prentice, R.~L. (1986).
\newblock A case-cohort design for epidemiologic cohort studies and disease prevention trials.
\newblock {\em Biometrika}, 73(1):1--11.

\bibitem[S{\"a}rndal, 2007]{sarndal2007calibration}
S{\"a}rndal, C.-E. (2007).
\newblock The calibration approach in survey theory and practice.
\newblock {\em Survey methodology}, 33(2):99--119.

\bibitem[Shin et~al., 2020]{shin2020weight}
Shin, Y.~E., Pfeiffer, R.~M., Graubard, B.~I., and Gail, M.~H. (2020).
\newblock Weight calibration to improve the efficiency of pure risk estimates from case-control samples nested in a cohort.
\newblock {\em Biometrics}, 76(4):1087--1097.

\bibitem[Thompson et~al., 2012]{thompson2012icc}
Thompson, D.~M., Fernald, D.~H., and Mold, J.~W. (2012).
\newblock Intraclass correlation coefficients typical of cluster-randomized studies: Estimates from the robert wood johnson prescription for health projects.
\newblock {\em The Annals of Family Medicine}, 10(3):235--240.

\bibitem[Wang, 2025]{wang2025using}
Wang, L. (2025).
\newblock Using model-assisted calibration methods to improve efficiency of regression analyses using two-phase samples or pooled samples under complex survey designs.
\newblock {\em Biometrics}, 81(3):ujaf092.

\bibitem[Wang et~al., 2025]{wang2025data}
Wang, L., Li, Y., Graubard, B.~I., and Katki, H.~A. (2025).
\newblock Data-integration with pseudoweights and survey-calibration: application to developing us-representative lung cancer risk models for use in screening.
\newblock {\em Journal of the Royal Statistical Society Series A: Statistics in Society}, 188(1):119--139.

\bibitem[Wu and Sitter, 2001]{wu2001model}
Wu, C. and Sitter, R.~R. (2001).
\newblock A model-calibration approach to using complete auxiliary information from survey data.
\newblock {\em Journal of the American Statistical Association}, 96(453):185--193.

\bibitem[Zhang et~al., 2020]{zhang2020generalized}
Zhang, H., Deng, L., Schiffman, M., Qin, J., and Yu, K. (2020).
\newblock Generalized integration model for improved statistical inference by leveraging external summary data.
\newblock {\em Biometrika}, 107(3):689--703.

\bibitem[Zheng et~al., 2022]{zheng2022risk}
Zheng, J., Zheng, Y., and Hsu, L. (2022).
\newblock Risk projection for time-to-event outcome leveraging summary statistics with source individual-level data.
\newblock {\em Journal of the American Statistical Association}, 117(540):2043--2055.

\end{thebibliography}

\end{document}


\begin{center}
{\LARGE \textbf{Supplementary Materials}}\\[1em]
{\large for}\\[0.5em]
{\large \textit{Improving Efficiency of Regression
Analyses by Integrating Data from
Population-Representative Surveys: A
Model-Assisted Calibration Approach}}\\[0.5em]
\textit{By Yanhao Lu and Lingxiao Wang}
\end{center}


\section{Proofs of Consistency of $\widehat{\pmb{\beta}}_{\mathrm{p.clb}}$ and $\widehat{\pmb{\beta}}_{\mathrm{e.clb}}$}

In this Supplementary Material, we provide the theoretical justification for the design consistency of
the pooled-sample calibration estimator $\widehat{\pmb{\beta}}_{\mathrm{p.clb}}$
(Section~3) and the external-sample calibration estimator $\widehat{\pmb{\beta}}_{\mathrm{e.clb}}$
(Section~4) under the combined use of the internal probability sample $s_1$ and the external
probability sample $s_2$.
Throughout this appendix, we adopt the notation and setup described in Section~2 of the main paper.

\subsection{Regularity Conditions}

Let $\{FP^{(k)}\}$ denote a sequence of finite populations of size $N^{(k)}$
from which the internal sample $s_1^{(k)}$ of size $n_1^{(k)}$
and the external sample $s_2^{(k)}$ of size $n_2^{(k)}$ are independently selected under
stratified multistage sampling designs.
We consider the asymptotic framework
\[
N^{(k)} \to \infty,\qquad
n_1^{(k)}\to\infty,\qquad
n_2^{(k)}\to\infty,\qquad
n_1^{(k)}/N^{(k)}=O(1),\,
n_2^{(k)}/N^{(k)}=O(1).
\]

\noindent We assume the following regularity conditions:

\begin{itemize}
\item[\textbf{C1.}]
Design consistency of the FP means: for any function $g_i$ of $(y_i,\pmb{x}_i)$ with finite expectation,
\[
N^{-1}\sum_{i\in s_d} w_i^{(s_d)} g_i
=
N^{-1}\sum_{i\in FP} g_i + O_p(n_d^{-1/2}),
\qquad d=1,2.
\]

\item[\textbf{C2.}]
Inclusion probabilities for $i\in \mathcal{U}$ satisfy
\[
0 < c_1 < \pi_i^{(s_d)} < c_2 < \infty,\qquad d=1,2.
\]
where $c_1, c_2\in R$
\item[\textbf{C3.}]
The covariates and responses satisfy
\[
N^{-1}\sum_{i\in FP} \|\pmb{x}_i\|^3 = O(1),\qquad
N^{-1}\sum_{i\in FP} \|\pmb{x}^*_i\|^3 = O(1),\qquad
N^{-1}\sum_{i\in FP} y_i^2 = O(1).
\]

\item[\textbf{C4.}]
The Fisher information matrix for the outcome model is positive definite (PD). For example, for a logistic regression model:
\[
N^{-1}\sum_{i\in FP} p_i(1-p_i)\pmb{x}_i\pmb{x}_i^\top \;\text{ is PD}.
\]

\item[\textbf{C5.}]
The Fisher information matrix for the surrogate model information is PD. For example, for a logistic regression model,
$N^{-1}\sum_{i\in FP} p_i^*(1-p_i^*)\pmb{x}^*_i\pmb{x}_i^{*\top}$
is positive definite.

\item[\textbf{C6.}]
The influence function of $\widehat{\pmb{\beta}}^*$
under the surrogate model is well-defined and satisfies
\[
N^{-1}\sum_{i\in FP} {\Delta}^*_i ({\Delta}^*_i)^\top\;\text{is PD}.
\]

\end{itemize}

\noindent Under Conditions C1--C5, $s_d$ provides 
are design-consistent for $\pmb{\beta}_{FP}$ and $\pmb{\beta}^*_{FP}$ if $\{(y_i, \pmb{x}_i, x_{2,i}^*), i\in s_d\}$ is fully observed.

\subsection{Influence Function of the Internal Estimator $\widehat{\pmb{\beta}}_{s_1}$}

For concreteness, we focus on the logistic regression model as an example. The FP parameter for the outcome model is defined by the solution of the FP estimation equation as below
\[
\pmb{U}_{FP}(\pmb{\beta})
= \sum_{i\in FP} \pmb{u}_i(\pmb{\beta})
= \sum_{i\in FP} (y_i - p_i)\pmb{x}_i = \pmb{0},
\quad
p_i=\operatorname{expit}(\pmb{x}_i^\top \pmb{\beta}).
\]

\noindent The internal estimator $\widehat{\pmb{\beta}}_{s_1}$
solves the sample-weighted equation
\begin{equation}
\widehat{\pmb{U}}^{(s_1)}(\pmb{\beta})
= \sum_{i\in s_1} w_i^{(s_1)}\pmb{u}_i(\pmb{\beta})=\pmb{0}.
\label{eq:A1}
\end{equation}
\noindent Under standard regularity conditions C1-C4 , 
a first-order Taylor expansion around $\pmb{\beta}_{FP}$ yields
\[
\widehat{\pmb{\beta}}_{s_1}-\pmb{\beta}_{FP}
=
\pmb{U}_{\pmb{\beta}}^{-1}
\sum_{i\in s_1}w_i^{(s_1)}\pmb{u}_i(\pmb{\beta}_{FP})
+ o_p(n_1^{-1/2}),
\]
where
\[
\pmb{U}_{\pmb{\beta}}
= \frac{\partial \pmb{U}_{FP}(\pmb{\beta})}{\partial\pmb{\beta}^\top}
\Big|_{\pmb{\beta}=\pmb{\beta}_{FP}}
= \sum_{i\in FP} p_i(1-p_i)\pmb{x}_i\pmb{x}_i^\top.
\]

Define the influence function
\[
\pmb{\Delta}_i = \pmb{U}_{\pmb{\beta}}^{-1}\pmb{u}_i(\pmb{\beta}_{FP}),\qquad
\widehat{\pmb{\Delta}}_i =
\widehat{\pmb{U}}_{\pmb{\beta}}^{-1}\pmb{u}_i(\widehat{\pmb{\beta}}_{s_1}).
\]

Then
\begin{equation}
\widehat{\pmb{\beta}}_{s_1}-\pmb{\beta}_{FP}
=
\sum_{i\in s_1} w_i^{(s_1)}\pmb{\Delta}_i
+ o_p(n_1^{-1/2}).
\label{eq:A2}
\end{equation}

\subsection{Influence Function of the Surrogate Estimators (Pooled-Sample)}

Define the surrogate score
\[
\pmb{u}_i^*(\pmb{\beta}^*)=(y_i-p_i^*)\pmb{x}_i^*,\qquad
p_i^*=\operatorname{expit}(\pmb{x}_i^{*\top}\pmb{\beta}^*).
\]

Let the combined sample be $S=s_1\cup s_2$ (with independent samples), with pooled weights $w_i^{(S)}$
defined as in Section~3 of the main paper. The pooled-sample surrogate estimator
$\widehat{\pmb{\beta}}^*_S$ solves
\begin{equation}
\widehat{\pmb{U}}^{*(S)}(\pmb{\beta}^*)
=\sum_{i\in S} w_i^{(S)} \pmb{u}_i^*(\pmb{\beta}^*)=\pmb{0}.
\label{eq:A3}
\end{equation}

\noindent Under standard regularity conditions C1-C3 and C5, a first-order Taylor expansion yields
\[
\widehat{\pmb{\beta}}^*_S-\pmb{\beta}^*_{FP}
=
\sum_{i\in S} w_i^{(S)}\pmb{\Delta}_i^*
+ o_p(n^{-1/2}),
\]
with
\[
\pmb{\Delta}_i^*=(\pmb{U}_{\pmb{\beta}^*})^{-1}\pmb{u}_i^*(\pmb{\beta}^*_{FP}),\qquad
\pmb{U}_{\pmb{\beta}^*}=\sum_{i\in FP}p_i^*(1-p_i^*)\pmb{x}_i^*\pmb{x}_i^{*\top}.
\]

\noindent Similarly, for the internal-only surrogate estimator:
\begin{equation}
\widehat{\pmb{\beta}}^*_{s_1}-\pmb{\beta}^*_{FP}
=
\sum_{i\in s_1} w_i^{(s_1)}\pmb{\Delta}_i^*
+ o_p(n_1^{-1/2}).
\label{eq:A4}
\end{equation}

\subsection{Calibration Equation and First-Order Expansion (Pooled-Sample)}

For the GREG-type pooled-sample calibration, the calibration factors take the linear form
\[
F_i(\pmb{\eta})=1+\pmb{\eta}^\top \pmb{\Delta}_i^*,\qquad i\in s_1,
\]
where $\pmb{\eta}$ solves the moment-matching equation
\begin{equation}
\sum_{i\in s_1} w_i^{(s_1)}F_i(\pmb{\eta})\pmb{\Delta}_i^*
=
\sum_{i\in S} w_i^{(S)} \pmb{\Delta}_i^*.
\label{eq:A5}
\end{equation}

\noindent Define
\[
\pmb{V}=\sum_{i\in s_1} w_i^{(s_1)}\pmb{\Delta}_i^*\pmb{\Delta}_i^{*\top},\qquad
\pmb{T}^*_{s_1}=\sum_{i\in s_1} w_i^{(s_1)}\pmb{\Delta}_i^*,\qquad
\pmb{T}^*_{S}=\sum_{i\in S} w_i^{(S)} \pmb{\Delta}_i^*.
\]

\noindent Under standard regularity conditions C1-C3, C5-C6, the first-order Taylor expansion of \eqref{eq:A5} around $\pmb{\eta}=\pmb{0}$, we obtain
\begin{equation}
\widehat{\pmb{\eta}}
=
\pmb{V}^{-1}(\pmb{T}^*_{S}-\pmb{T}^*_{s_1})
+ o_p(n^{-1/2}).
\label{eq:A6}
\end{equation}

\subsection{Proof of Consistency of $\widehat{\pmb{\beta}}_{\mathrm{p.clb}}$ (Theorem~3.1)}
\label{subsec:proof-pclb}

The pooled-sample calibration estimator solves
\[
\widehat{\pmb{U}}_{\mathrm{p.clb}}(\pmb{\beta})
=
\sum_{i\in s_1} w_i^{(s_1)}F_i(\widehat{\pmb{\eta}})\pmb{u}_i(\pmb{\beta})=\pmb{0}.
\]

\noindent Define
\[
\pmb{\Theta}
=
\left(
\sum_{i\in s_1} w_i^{(s_1)}\pmb{\Delta}_i^*\pmb{\Delta}_i^{*\top}
\right)^{-1}
\left(
\sum_{i\in s_1} w_i^{(s_1)}\pmb{\Delta}_i^*\pmb{\Delta}_i^{\top}
\right).
\]

\noindent Using the expansion in \eqref{eq:A2}, we have
\[
\widehat{\pmb{\beta}}_{\mathrm{p.clb}}
- \pmb{\beta}_{FP}
=
\sum_{i\in s_1} w_i^{(s_1)}F_i(\widehat{\pmb{\eta}})\pmb{\Delta}_i
+ o_p(n_1^{-1/2}).
\]

\noindent Substituting $F_i(\widehat{\pmb{\eta}})=1+\widehat{\pmb{\eta}}^\top\pmb{\Delta}_i^*$ gives
\[
\sum_{i\in s_1} w_i^{(s_1)}F_i(\widehat{\pmb{\eta}})\pmb{\Delta}_i
=
\sum_{i\in s_1} w_i^{(s_1)}\pmb{\Delta}_i
+
\left(
\sum_{i\in s_1} w_i^{(s_1)}\pmb{\Delta}_i\pmb{\Delta}_i^{*\top}
\right)\widehat{\pmb{\eta}}.
\]

\noindent Using \eqref{eq:A2} and \eqref{eq:A4}, we have
\[
\sum_{i\in s_1} w_i^{(s_1)}\pmb{\Delta}_i
=
\widehat{\pmb{\beta}}_{s_1}-\pmb{\beta}_{FP}
+ o_p(n_1^{-1/2}),
\]
and from \eqref{eq:A6},
\[
\widehat{\pmb{\eta}}
=
\pmb{V}^{-1}
\left(
\pmb{T}_S^*-\pmb{T}_{s_1}^*
\right)
+ o_p(n^{-1/2}),
\]
where $\pmb{V}=\sum_{i\in s_1} w_i^{(s_1)}\pmb{\Delta}_i^*\pmb{\Delta}_i^{*\top}$.\\

\noindent Therefore,
\[
\left(
\sum_{i\in s_1} w_i^{(s_1)}\pmb{\Delta}_i\pmb{\Delta}_i^{*\top}
\right)\widehat{\pmb{\eta}}
=
\pmb{\Theta}^{\top}
\left(
\pmb{T}_S^*-\pmb{T}_{s_1}^*
\right)
+ o_p(n^{-1/2}).
\]

\noindent Using \eqref{eq:A3} and \eqref{eq:A4}, we further obtain
\[
\pmb{T}_S^*-\pmb{T}_{s_1}^*
=
\widehat{\pmb{\beta}}_S^*-\widehat{\pmb{\beta}}_{s_1}^*
+ o_p(n^{-1/2}).
\]

\noindent Combining the above results yields
\[
\widehat{\pmb{\beta}}_{\mathrm{p.clb}}
    =
    \widehat{\pmb{\beta}}_{s_1}
    +
    \pmb{\Theta}^{\top}
    \left(
      \widehat{\pmb{\beta}}^{*}_S-\widehat{\pmb{\beta}}^{*}_{s_1}
    \right)
    + o_p(n^{-1/2}).
\]

\noindent Consequently,
\[
\widehat{\pmb{\beta}}_{\mathrm{p.clb}}
\overset{p}{\longrightarrow}
\pmb{\beta}_{FP}.
\]

\noindent This completes the proof of Theorem~3.1.

\subsection{Equivalence of Score- and Influence-Function Calibration}

\begin{lemma}[Equivalence of Score- and Influence-Function Calibration]
\label{lem:score-if-equiv}
Suppose the surrogate score Jacobian $\pmb{U}_{\pmb{\beta}^*}^*$ is nonsingular.
Let the surrogate influence function be defined as
\[
\pmb{\Delta}_i^* = (\pmb{U}_{\pmb{\beta}^*}^*)^{-1} \pmb{u}_i^*,
\qquad
\pmb{u}_i^* = (y_i - p_i^*) \pmb{x}_i^*.
\]
Then calibration based on auxiliary variables $\{\pmb{\Delta}_i^*\}$ is algebraically
equivalent to calibration based directly on surrogate score contributions
$\{\pmb{u}_i^*\}$ in the sense that both approaches yield identical calibrated weights
and hence the same calibrated estimator.
\end{lemma}

\begin{proof}
By definition of the influence function,
\[
\pmb{u}_i^* = \pmb{U}_{\pmb{\beta}^*}^* \pmb{\Delta}_i^*,
\]
where $\pmb{U}_{\pmb{\beta}^*}^*$ is a nonsingular matrix that does not
depend on the unit index $i$.

\noindent Consider first the calibration equations based on influence-function auxiliary
variables. Under a chi-square distance measure, the calibration equation takes
the form
\begin{equation}
\sum_{i \in s_1} w_i^{(s_1)} \{1 + \pmb{\eta}^\top \pmb{\Delta}_i^*\} \pmb{\Delta}_i^*
=
\sum_{i \in S} w_i^{(S)} \pmb{\Delta}_i^* .
\label{eq:A-IF-calib}
\end{equation}

\noindent Substituting $\pmb{\Delta}_i^* = (\pmb{U}_{\pmb{\beta}^*}^*)^{-1} \pmb{u}_i^*$ into \eqref{eq:A-IF-calib} gives
\[
(\pmb{U}_{\pmb{\beta}^*}^*)^{-1}
\left[
\sum_{i \in s_1} w_i^{(s_1)} \{1 + \pmb{\eta}^\top (\pmb{U}_{\pmb{\beta}^*}^*)^{-1} \pmb{u}_i^*\}\pmb{u}_i^*
\right]
=
(\pmb{U}_{\pmb{\beta}^*}^*)^{-1}
\sum_{i \in S} w_i^{(S)} \pmb{u}_i^*.
\]

\noindent Multiplying both sides by $\pmb{U}_{\pmb{\beta}^*}^*$ yields the equivalent score-based calibration equation
\begin{equation}
\sum_{i \in s_1} w_i^{(s_1)} \{1 + \tilde{\pmb{\eta}}^\top \pmb{u}_i^*\}\pmb{u}_i^*
=
\sum_{i \in S} w_i^{(S)} \pmb{u}_i^*,
\label{eq:A-score-calib}
\end{equation}
where $\tilde{\pmb{\eta}} = (\pmb{U}_{\pmb{\beta}^*}^*)^{-1} \pmb{\eta}$.\\

\noindent Since $\pmb{U}_{\pmb{\beta}^*}^*$ is nonsingular, the mapping
$\pmb{\eta} \leftrightarrow \tilde{\pmb{\eta}}$ is one-to-one, and the resulting calibration
factors satisfy
\[
F_i = 1 + \pmb{\eta}^\top \pmb{\Delta}_i^*
= 1 + \tilde{\pmb{\eta}}^\top \pmb{u}_i^*.
\]
Therefore, both formulations yield identical calibrated weights and the same
calibrated estimator.
\end{proof}
\medskip
\noindent
\textbf{Remark.}
Although calibration can be expressed using surrogate influence functions
$\{\pmb{\Delta}_i^{*}, i\in S\}$, Lemma~\ref{lem:score-if-equiv} shows that, under generalized linear models,
this is algebraically equivalent to calibrating directly on surrogate score
contributions $\{\pmb{u}_i^{*}, i\in S\}$.
In particular, the Jacobian matrix $\pmb{U}_{\pmb{\beta}^{*}}^{*}$ cancels out of the
calibration equations.

From a practical perspective, this equivalence implies that it is unnecessary
to construct plug-in influence functions involving the inverse Jacobian
$\widehat{\pmb{U}}_{\pmb{\beta}^*}^{-1}$.
Such plug-in estimators may amplify estimation noise through matrix inversion,
especially when the internal sample size is moderate.
Implementing calibration directly with score contributions therefore achieves
the same first-order target while improving numerical stability in finite samples.

\subsection{Influence Function of the Surrogate Estimators (External-Sample)}

Define the surrogate score
\[
\pmb{u}_i^*(\pmb{\beta}^*)=(y_i-p_i^*)\pmb{x}_i^*,\qquad
p_i^*=\operatorname{expit}(\pmb{x}_i^{*\top}\pmb{\beta}^*).
\]

\noindent The external surrogate estimator
$\widehat{\pmb{\beta}}^*_{s_2}$ solves
\begin{equation}
\widehat{\pmb{U}}^{*(s_2)}(\pmb{\beta}^*)
=\sum_{i\in s_2} w_i^{(s_2)} \pmb{u}_i^*(\pmb{\beta}^*)=\pmb{0}.
\label{eq:surr_s2}
\end{equation}

\noindent Under standard regularity conditions C1-C3, and C5, a first-order Taylor expansion yields
\[
\widehat{\pmb{\beta}}^*_{s_2}-\pmb{\beta}^*_{FP}
=
\sum_{i\in s_2} w_i^{(s_2)}\pmb{\Delta}_i^*
+ o_p(n_2^{-1/2}),
\]
with
\[
\pmb{\Delta}_i^*=(\pmb{U}_{\pmb{\beta}^*})^{-1}\pmb{u}_i^*(\pmb{\beta}^*_{FP}),\qquad
\pmb{U}_{\pmb{\beta}^*}=\sum_{i\in FP}p_i^*(1-p_i^*)\pmb{x}_i^*\pmb{x}_i^{*\top}.
\]

\noindent The same expansion \eqref{eq:A4} holds for the internal-only surrogate estimator:
\[
\widehat{\pmb{\beta}}^*_{s_1}-\pmb{\beta}^*_{FP}
=
\sum_{i\in s_1} w_i^{(s_1)}\pmb{\Delta}_i^*
+ o_p(n_1^{-1/2}).
\]

\subsection{Calibration Equation and First-Order Expansion (External-Sample)}

For the GREG-type external-sample calibration, the calibration factors take the linear form
\[
F_i(\pmb{\eta})=1+\pmb{\eta}^\top \pmb{u}_i^*,\qquad i\in s_1,
\]
where $\pmb{\eta}$ solves the moment-matching equation
\begin{equation}
\sum_{i\in s_1} w_i^{(s_1)}F_i(\pmb{\eta})\pmb{u}_i^*
=
\sum_{i\in s_2} w_i^{(s_2)} \pmb{u}_i^*.
\label{eq:A8}
\end{equation}

\textbf{}Define
\[
\pmb{V}=\sum_{i\in s_1} w_i^{(s_1)}\pmb{u}_i^*\pmb{u}_i^{*\top},\qquad
\pmb{T}^*_{s_1}=\sum_{i\in s_1} w_i^{(s_1)}\pmb{u}_i^*,\qquad
\pmb{T}^*_{s_2}=\sum_{i\in s_2} w_i^{(s_2)}\pmb{u}_i^*.
\]

\textbf{}Under standard regularity conditions C1-3, C5-C6, and a first-order Taylor expansion of \eqref{eq:A8} around $\pmb{\eta}=\pmb{0}$, we obtain
\begin{equation}
\widehat{\pmb{\eta}}
=
\pmb{V}^{-1}(\pmb{T}^*_{s_2}-\pmb{T}^*_{s_1})
+ o_p(n^{-1/2}).
\label{eq:A9}
\end{equation}

\subsection{Proof of Consistency of $\widehat{\pmb{\beta}}_{\mathrm{e.clb}}$ (Theorem~4.2)}

The external-sample calibration estimator solves
\[
\widehat{\pmb{U}}_{\mathrm{e.clb}}(\pmb{\beta})
=
\sum_{i\in s_1} w_i^{(s_1)}F_i(\widehat{\pmb{\eta}})\pmb{u}_i(\pmb{\beta})=\pmb{0}.
\]

\noindent Here $\pmb{\Theta}$ denotes the matrix defined in Section~\ref{subsec:proof-pclb}.

\noindent Using the influence-function expansion \eqref{eq:A2}, we have
\[
\widehat{\pmb{\beta}}_{\mathrm{e.clb}}
- \pmb{\beta}_{FP}
=
\sum_{i\in s_1} w_i^{(s_1)}F_i(\widehat{\pmb{\eta}})\pmb{\Delta}_i
+ o_p(n_1^{-1/2}).
\]

\noindent Substituting the linear calibration factor $F_i(\widehat{\pmb{\eta}})=1+\widehat{\pmb{\eta}}^\top \pmb{u}_i^*$ yields
\[
\sum_{i\in s_1} w_i^{(s_1)}F_i(\widehat{\pmb{\eta}})\pmb{\Delta}_i
=
\sum_{i\in s_1} w_i^{(s_1)}\pmb{\Delta}_i
+
\sum_{i\in s_1} w_i^{(s_1)}\widehat{\pmb{\eta}}^{\top}\pmb{u}_i^*\,\pmb{\Delta}_i.
\]

\noindent Using the first-order expansion of $\widehat{\pmb{\eta}}$ in \eqref{eq:A9}, substituting
$\pmb{u}_i^*=\pmb{U}_{\pmb{\beta}^*}\pmb{\Delta}_i^*$ and recognising that
$\left(\sum_{i\in s_1}w_i^{(s_1)}\pmb{\Delta}_i\pmb{u}_i^{*\top}\right)\pmb{V}^{-1}
=\pmb{\Theta}^\top\pmb{U}_{\pmb{\beta}^*}^{-1}$,
and using $\pmb{U}_{\pmb{\beta}^*}^{-1}\pmb{T}^*_{s_k}
=\widehat{\pmb{\beta}}^*_{s_k}-\pmb{\beta}^*_{FP}+o_p(n_k^{-1/2})$
(from \eqref{eq:A4} for $k=1$ and its analogue for $k=2$), we obtain
\[
\sum_{i\in s_1} w_i^{(s_1)}F_i(\widehat{\pmb{\eta}})\pmb{\Delta}_i
=
\widehat{\pmb{\beta}}_{s_1}-\pmb{\beta}_{FP}
+
\pmb{\Theta}^\top
\left(
\widehat{\pmb{\beta}}^*_{s_2}-\widehat{\pmb{\beta}}^*_{s_1}
\right)
+ o_p(n_1^{-1/2}).
\]

\noindent Therefore,
\[
\widehat{\pmb{\beta}}_{\mathrm{e.clb}}
= \widehat{\pmb{\beta}}_{s_1}
+
\pmb{\Theta}^\top
\left(
\widehat{\pmb{\beta}}^{*}_{s_2}- \widehat{\pmb{\beta}}^{*}_{s_1}
\right) + o_p(n_1^{-1/2}).
\]

\noindent Since $\widehat{\pmb{\beta}}_{s_1}\overset{p}{\to}\pmb{\beta}_{FP}$ and
$\widehat{\pmb{\beta}}^*_{s_2}, \widehat{\pmb{\beta}}^*_{s_1}\overset{p}{\to}\pmb{\beta}^*_{FP}$,
the second term converges to $\pmb{0}$, and thus
\[
\widehat{\pmb{\beta}}_{\mathrm{e.clb}}
\overset{p}{\longrightarrow}
\pmb{\beta}_{FP}.
\]

This completes the proof of Theorem~4.2.

\subsection{Why Not Use Sample Plug-in Influence Functions}

In this subsection, we explain why it is preferable in practice to avoid calibration based on
sample plug-in influence functions, even though such a construction remains asymptotically valid.

Recall that the finite-population influence functions are
\[
\pmb{\Delta}_i
=
\pmb{U}_{\pmb{\beta}}^{-1}\pmb{u}_i(\pmb{\beta}_{FP}),
\qquad
\pmb{\Delta}_i^*
=
\left(\pmb{U}_{\pmb{\beta}^*}\right)^{-1}
\pmb{u}_i^*(\pmb{\beta}_{FP}^*).
\]

\noindent A natural plug-in version replaces the unknown finite-population quantities by their sample
estimators:
\[
\widehat{\pmb{\Delta}}_i
=
\widehat{\pmb{U}}_{\pmb{\beta}}^{-1}
\pmb{u}_i(\widehat{\pmb{\beta}}_{s_1}),
\qquad
\widehat{\pmb{\Delta}}_i^*
=
\left(\widehat{\pmb{U}}_{\pmb{\beta}^*}^{*(s_2)}\right)^{-1}
\pmb{u}_i^*(\widehat{\pmb{\beta}}_{s_2}^*).
\]

\noindent Under Conditions C1--C6 and standard smoothness arguments,
\[
\widehat{\pmb{U}}_{\pmb{\beta}}
=
\pmb{U}_{\pmb{\beta}} + O_p(n_1^{-1/2}),
\qquad
\widehat{\pmb{\beta}}_{s_1}
=
\pmb{\beta}_{FP} + O_p(n_1^{-1/2}),
\]
and similarly
\[
\widehat{\pmb{U}}_{\pmb{\beta}^*}^{*(s_2)}
=
\pmb{U}_{\pmb{\beta}^*} + O_p(n_2^{-1/2}),
\qquad
\widehat{\pmb{\beta}}_{s_2}^*
=
\pmb{\beta}_{FP}^* + O_p(n_2^{-1/2}).
\]

\noindent By a first-order matrix expansion,
\[
\widehat{\pmb{U}}_{\pmb{\beta}}^{-1}
=
\pmb{U}_{\pmb{\beta}}^{-1} + O_p(n_1^{-1/2}),
\qquad
\left(\widehat{\pmb{U}}_{\pmb{\beta}^*}^{*(s_2)}\right)^{-1}
=
\pmb{U}_{\pmb{\beta}^*}^{-1} + O_p(n_2^{-1/2}),
\]
so that
\begin{equation}
\widehat{\pmb{\Delta}}_i
=
\pmb{\Delta}_i + O_p(n_1^{-1/2}),
\qquad
\widehat{\pmb{\Delta}}_i^*
=
\pmb{\Delta}_i^* + O_p(n_2^{-1/2}).
\label{eq:A7}
\end{equation}

\noindent Now consider the plug-in versions of the calibration totals and calibration matrix:
\[
\widehat{\pmb{T}}_{s_1}^*
=
\sum_{i\in s_1} w_i^{(s_1)} \widehat{\pmb{\Delta}}_i^*,
\qquad
\widehat{\pmb{V}}
=
\sum_{i\in s_1} w_i^{(s_1)}
\widehat{\pmb{\Delta}}_i^* \widehat{\pmb{\Delta}}_i^{*\top}.
\]

\noindent Using \eqref{eq:A7}, we obtain
\[
\widehat{\pmb{T}}_{s_1}^*
=
\sum_{i\in s_1} w_i^{(s_1)} \pmb{\Delta}_i^*
+ O_p(n_1^{-1/2}),
\]
and
\[
\widehat{\pmb{V}}
=
\sum_{i\in s_1} w_i^{(s_1)}
\pmb{\Delta}_i^* \pmb{\Delta}_i^{*\top}
+ O_p(n_1^{-1/2}).
\]

\noindent Therefore, calibration based on $\widehat{\pmb{\Delta}}_i^*$ yields the same first-order limit
as calibration based on $\{\pmb{\Delta}_i^*, i\in S\}$, and thus does not alter first-order consistency.

However, unlike calibration based directly on surrogate score contributions
$\{\pmb{u}_i^*, i\in S\}$, the plug-in construction introduces additional randomness through the
estimated inverse Jacobian matrices
$\widehat{\pmb{U}}_{\pmb{\beta}}^{-1}$ and
$\left(\widehat{\pmb{U}}_{\pmb{\beta}^*}^{*(s_2)}\right)^{-1}$.
This additional $O_p(n^{-1/2})$ perturbation enters both the calibration totals and the
matrix $\pmb{V}$, and therefore propagates into the calibration adjustment
$\widehat{\pmb{\eta}}$ and the resulting calibrated estimator.

Consequently, although plug-in influence-function calibration is asymptotically valid, it may
exhibit larger finite-sample variability than score-based calibration.

By contrast, Lemma~\ref{lem:score-if-equiv} shows that when calibration is formulated directly
in terms of surrogate score contributions, the Jacobian cancels out algebraically.
Hence score-based calibration attains the same first-order target while avoiding the extra
sampling variability induced by estimating and inverting Jacobian matrices.
This provides theoretical justification for using score functions rather than sample plug-in
influence functions in implementation.

\section{Taylor--Linearization Variance of 
$\widehat{\pmb{\beta}}_{\mathrm{p.clb}}$ and 
$\widehat{\pmb{\beta}}_{\mathrm{e.clb}}$}
\label{app:var_fullproof}

\subsection{General stacked estimating equation system}

Let $\pmb{\theta}=(\pmb{\beta}^\top,\pmb{\eta}^\top)^\top$ denote the stacked parameter vector,
where $\pmb{\beta}$ is the parameter of interest and $\pmb{\eta}$ is the calibration parameter.
Both pooled-sample and external-sample calibration estimators can be written as solutions to
a stacked estimating equation system:
\begin{equation}
\pmb{\Psi}(\pmb{\theta})=
\begin{pmatrix}
\pmb{U}(\pmb{\beta},\pmb{\eta})\\
\pmb{Q}(\pmb{\beta},\pmb{\eta})
\end{pmatrix}
=\pmb{0},
\label{eq:app_stack}
\end{equation}
where $\pmb{U}$ is the weighted score equation for $\pmb{\beta}$ and $\pmb{Q}$ is the calibration equation.\\

Let $\pmb{\theta}_0=(\pmb{\beta}_{FP}^\top,\pmb{\eta}_{FP}^\top)^\top$ denote the finite-population solution, 
i.e., the solution to $\pmb{\Psi}_{FP}(\pmb{\theta})=\pmb{0}$ defined at the finite-population level.

\subsection{First-order Taylor linearization}

We expand $\pmb{\Psi}(\pmb{\theta})$ around $\pmb{\theta}_0$:
\begin{equation}
\pmb{\Psi}(\widehat{\pmb{\theta}})
=
\pmb{\Psi}(\pmb{\theta}_0)
+
\pmb{\psi}(\pmb{\theta}_0)(\widehat{\pmb{\theta}}-\pmb{\theta}_0)
+
o_p(n^{-1/2}),
\label{eq:app_taylor}
\end{equation}
where $\pmb{\psi}(\pmb{\theta})=\partial \pmb{\Psi}(\pmb{\theta})/\partial \pmb{\theta}$
denotes the Jacobian matrix of $\pmb{\Psi}(\pmb{\theta})$.\\

\noindent Since $\pmb{\Psi}(\widehat{\pmb{\theta}})=\pmb{0}$ and $\pmb{\psi}(\pmb{\theta}_0)$ is nonsingular, rearranging \eqref{eq:app_taylor} gives
\begin{equation}
\widehat{\pmb{\theta}}-\pmb{\theta}_0
=
-\pmb{\psi}^{-1}(\pmb{\theta}_0)\pmb{\Psi}(\pmb{\theta}_0)
+o_p(n^{-1/2}).
\label{eq:app_linrep}
\end{equation}

\noindent Taking variance on both sides yields
\begin{equation}
\operatorname{Var}(\widehat{\pmb{\theta}})
=
\pmb{\psi}^{-1}(\pmb{\theta}_0)
\operatorname{Var}\{\pmb{\Psi}(\pmb{\theta}_0)\}
\pmb{\psi}^{-1}(\pmb{\theta}_0)^\top +o_p(n^{-1}),
\label{eq:app_var_theta}
\end{equation}
where the approximation holds to first order.

\subsection{Structure of the Jacobian matrix}

The Jacobian matrix has a block triangular structure:
\begin{equation}
\pmb{\Psi}(\pmb{\theta}_0)
=
E
\left[
\left.
\frac{\partial \pmb{\psi}(\pmb{\theta})}
{\partial \pmb{\theta}^{\top}}
\right|_{\pmb{\theta}=\pmb{\theta}_0}
\right]
=
\begin{pmatrix}
\pmb{U}_{\pmb{\beta}} & \pmb{U}_{\pmb{\eta}}\\
\pmb{0} & \pmb{Q}_{\pmb{\eta}}
\end{pmatrix},
\label{eq:app_jacobian}
\end{equation}
Here $E(\cdot)$ denotes expectation with respect to the sampling design,
and $\pmb{\theta}_0=(\pmb{\beta}_0^\top,\pmb{\eta}_0^\top)^\top$.
Then the blocks in the expected Jacobian are
\[
\pmb{U}_{\pmb{\beta}}
=
E
\left[
\left.
\frac{\partial \pmb{U}(\pmb{\beta},\pmb{\eta})}
{\partial \pmb{\beta}^{\top}}
\right|_{\pmb{\theta}=\pmb{\theta}_0}
\right]
=
-
\sum_{i\in FP}
F_i(\pmb{\eta}_0)
p_i(\pmb{\beta}_0)
\{1-p_i(\pmb{\beta}_0)\}
\pmb{x}_i\pmb{x}_i^\top,
\]
\[
\pmb{U}_{\pmb{\eta}}
=
E
\left[
\left.
\frac{\partial \pmb{U}(\pmb{\beta},\pmb{\eta})}
{\partial \pmb{\eta}^{\top}}
\right|_{\pmb{\theta}=\pmb{\theta}_0}
\right]
=
\sum_{i\in FP}
\pmb{u}_i(\pmb{\beta}_0)\pmb{u}_i^{*\top},
\]
and
\[
\pmb{Q}_{\pmb{\eta}}
=
E
\left[
\left.
\frac{\partial \pmb{Q}(\pmb{\eta})}
{\partial \pmb{\eta}^{\top}}
\right|_{\pmb{\eta}=\pmb{\eta}_0}
\right]
=
\sum_{i\in FP}
\pmb{u}_i^*\pmb{u}_i^{*\top}.
\]
In practice, these matrices are estimated by
\[
\widehat{\pmb{U}}_{\pmb{\beta}}
=
-
\sum_{i\in s_1}
w_i^{(s_1)}
\widehat{F}_i
\widehat{p}_i(1-\widehat{p}_i)
\pmb{x}_i\pmb{x}_i^\top,
\]
\[
\widehat{\pmb{U}}_{\pmb{\eta}}
=
\sum_{i\in s_1}
w_i^{(s_1)}
\widehat{\pmb{u}}_i
\widehat{\pmb{u}}_i^{*\top},
\qquad
\widehat{\pmb{Q}}_{\pmb{\eta}}
=
\sum_{i\in s_1}
w_i^{(s_1)}
\widehat{\pmb{u}}_i^*
\widehat{\pmb{u}}_i^{*\top},
\]
where
\[
\widehat{F}_i=1+\widehat{\pmb{\eta}}^\top\widehat{\pmb{u}}_i^*,
\qquad
\widehat{p}_i=\operatorname{expit}(\pmb{x}_i^\top\widehat{\pmb{\beta}}_{s1}),
\qquad
\widehat{\pmb{u}}_i
=
\{y_i-\widehat{p}_i\}\pmb{x}_i,
\]
and
\[
\widehat{\pmb{u}}_i^*
=
\{y_i-p_i^*(\widehat{\pmb{\beta}}_{s_2}^*)\}\pmb{x}_i^*.
\]
The lower-left block is zero because the calibration equation
$\pmb{Q}(\pmb{\eta})$ does not depend on $\pmb{\beta}$ under the score-based formulation.

The lower-left block is zero because the calibration equation $\pmb{Q}(\pmb{\beta},\pmb{\eta})$
does not depend on $\pmb{\beta}$ under the score-based formulation.

Using block matrix inversion, we obtain
\begin{equation}
\pmb{\psi}^{-1}(\pmb{\theta}_0)
=
\begin{pmatrix}
\pmb{U}_{\pmb{\beta}}^{-1} & \pmb{b}\\
\pmb{0} & \pmb{Q}_{\pmb{\eta}}^{-1}
\end{pmatrix},
\quad
\pmb{b}=-\pmb{U}_{\pmb{\beta}}^{-1}\pmb{U}_{\pmb{\eta}}\pmb{Q}_{\pmb{\eta}}^{-1}.
\label{eq:app_inv_jacobian}
\end{equation}

\subsection{Variance decomposition for $\widehat{\pmb{\beta}}$}

Let
\begin{equation}
\operatorname{Var}\{\pmb{\Psi}(\pmb{\theta}_0)\}
=
\begin{pmatrix}
\pmb{V}_{11} & \pmb{V}_{12}\\
\pmb{V}_{12}^\top & \pmb{V}_{22}
\end{pmatrix},
\label{eq:app_varPsi}
\end{equation}
where $\pmb{V}_{21}=\pmb{V}_{12}^\top$.\\

\noindent Substituting \eqref{eq:app_inv_jacobian} into \eqref{eq:app_var_theta} and extracting the $\pmb{\beta}$ block gives
\begin{align}
\operatorname{Var}(\widehat{\pmb{\beta}})
&=
\pmb{U}_{\pmb{\beta}}^{-1}\pmb{V}_{11}\pmb{U}_{\pmb{\beta}}^{-1\top}
+
\pmb{U}_{\pmb{\beta}}^{-1}\pmb{V}_{12}\pmb{b}^\top
\nonumber\\
&\quad+
\pmb{b}\pmb{V}_{12}^\top\pmb{U}_{\pmb{\beta}}^{-1\top}
+
\pmb{b}\pmb{V}_{22}\pmb{b}^\top + o(n_1^{-1}),
\label{eq:app_var_beta_master}
\end{align}
where the approximation holds to first order.\\

The first term corresponds to the variance from the score equation, while the remaining terms capture
additional variability induced by calibration through the estimation of $\pmb{\eta}$.

\subsection{Pooled-sample calibration}

For pooled-sample calibration, variability arises from both $s_1$ and $s_2$. 
Under independence of $s_1$ and $s_2$ (Assumption~(A.2) of the main text),
\begin{equation}
\pmb{V}_{11}=\pmb{V}_{11}^{(s_1)},\quad
\pmb{V}_{12}=\pmb{V}_{12}^{(s_1)},\quad
\pmb{V}_{22}=\pmb{V}_{22}^{(s_1)}+\pmb{V}_{22}^{(s_2)}.
\label{eq:app_var_blocks_pooled}
\end{equation}

\noindent Substituting into \eqref{eq:app_var_beta_master} gives
\begin{align}
\operatorname{Var}(\widehat{\pmb{\beta}}_{\mathrm{p.clb}})
& = 
\pmb{U}_{\pmb{\beta}}^{-1}\pmb{V}_{11}^{(s_1)}\pmb{U}_{\pmb{\beta}}^{-1\top}
+
\pmb{U}_{\pmb{\beta}}^{-1}\pmb{V}_{12}^{(s_1)}\pmb{b}^\top
+
\pmb{b}\pmb{V}_{12}^{(s_1)\top}\pmb{U}_{\pmb{\beta}}^{-1\top}
\nonumber\\
&\quad+
\pmb{b}\pmb{V}_{22}^{(s_1)}\pmb{b}^\top
+
\pmb{b}\pmb{V}_{22}^{(s_2)}\pmb{b}^\top +  o(n_1^{-1}),
\label{eq:app_var_pooled}
\end{align}
where the approximation holds to first order.

The last term represents the additional variability contributed by the external sample through the calibration step.

\subsection{External-sample calibration}

When only summary information from $s_2$ is available, the contribution of $s_2$ enters through
$\widehat{\pmb{\beta}}^{*}_{s_2}$. By standard linearization,
\begin{equation}
\pmb{V}_{22}
=
\pmb{V}_{22}^{(s_1)}
+
\pmb{U}_{\pmb{\eta}}
\operatorname{Var}_{s_2}(\widehat{\pmb{\beta}}^{*})
\pmb{U}_{\pmb{\eta}}^\top.
\label{eq:app_var_blocks_external}
\end{equation}

\noindent Substituting into \eqref{eq:app_var_beta_master} yields
\begin{align}
\operatorname{Var}(\widehat{\pmb{\beta}}_{\mathrm{e.clb}})
&=
\pmb{U}_{\pmb{\beta}}^{-1}\pmb{V}_{11}^{(s_1)}\pmb{U}_{\pmb{\beta}}^{-1\top}
+
\pmb{U}_{\pmb{\beta}}^{-1}\pmb{V}_{12}^{(s_1)}\pmb{b}^\top
+
\pmb{b}\pmb{V}_{12}^{(s_1)\top}\pmb{U}_{\pmb{\beta}}^{-1\top}
\nonumber\\
&\quad+
\pmb{b}\pmb{V}_{22}^{(s_1)}\pmb{b}^\top
+
\pmb{b}
\Big(
\pmb{U}_{\pmb{\eta}}
\operatorname{Var}_{s_2}(\widehat{\pmb{\beta}}^{*})
\pmb{U}_{\pmb{\eta}}^\top
\Big)
\pmb{b}^\top +  o(n_1^{-1}),
\label{eq:app_var_external}
\end{align}

\noindent where the approximation holds to first order.\\

\noindent This expression shows that external-sample calibration propagates the uncertainty of the external estimator 
$\widehat{\pmb{\beta}}^{*}_{s_2}$ into the final estimator through 
$\operatorname{Var}_{s_2}(\widehat{\pmb{\beta}}^{*})$, without requiring access to microdata from $s_2$.

\bigskip
\noindent
This completes the Taylor--linearized variance derivation.

\section{Supplements for Simulation Studies}

\subsection{Details of the two-stage cluster sampling designs}

We generated both the internal sample $s_1$ and the external sample $s_2$
using two-stage cluster sampling designs from the same finite population. In
the simulation, the finite population contains $N=1{,}000{,}000$ individuals
and is divided into $M=10{,}000$ primary sampling units (PSUs), with
approximately equal PSU sizes.

For sample $s_k$, $k=1,2$, the first stage selects $m_k$ PSUs by simple random
sampling without replacement. Thus, the first-stage inclusion probability for
PSU $c$ is
\[
\pi_c^{(k)}=\frac{m_k}{M}.
\]
In the second stage, individuals within each selected PSU are sampled without
replacement using probability proportional to size (PPS) sampling. The PPS
size measure depends on covariates and the outcome, so that the resulting
sampling design is informative.

Specifically, for individual $i$ in sample $s_k$, we define the sampling size
measure
\[
q_{ki}
=
\exp\left\{
\gamma_{k1}x_{1,i}
+
\gamma_{k2}x_{2,i}
+
\gamma_{k3}y_i
+
\gamma_{k4}x_{1,i}y_i
+
\gamma_{k5}x_{2,i}y_i
\right\}.
\]
Here $q_{ki}$ is not itself the inclusion probability. Instead, it is the
relative PPS measure used to determine the chance that individual $i$ is
selected within a sampled PSU. Individuals with larger values of $q_{ki}$ have
larger second-stage selection probabilities.

The two samples use different size measures. In the simulation,
\[
\pmb{\gamma}_1
=
(0,\ 0.05,\ 0.10,\ 0,\ 0.01)^\top,
\qquad
\pmb{\gamma}_2
=
(0.05,\ 0,\ 0.10,\ 0.01,\ 0)^\top.
\]
Therefore, both sampling designs depend on the outcome $Y$, but the two
samples differ in how the covariates and outcome--covariate interactions enter
the PPS size measure.

Conditional on PSU $c$ being selected, approximately $n_k/m_k$ individuals are
sampled from that PSU using PPS sampling with size measure $q_{ki}$. Let
$n_{kc}$ denote the number of individuals sampled from PSU $c$ for sample
$s_k$. The conditional second-stage inclusion probability is approximately
\[
\pi_{i\mid c}^{(k)}
\approx
\frac{n_{kc}q_{ki}}{\sum_{j\in c}q_{kj}},
\qquad i\in c.
\]
Accordingly, the second-stage sampling weight used in the implementation is
\[
w_{i\mid c}^{(k)}
=
\frac{\sum_{j\in c}q_{kj}}{n_{kc}q_{ki}}.
\]

\noindent Combining the two stages, the overall inclusion probability for individual
$i\in FP$ in sample $s_k$ is approximately
\[
\pi_i^{(k)}
=
\pi_{c(i)}^{(k)}\pi_{i\mid c(i)}^{(k)}
\approx
\frac{m_k}{M}
\cdot
\frac{n_{k,c(i)}q_{ki}}{\sum_{j\in c(i)}q_{kj}},
\]
where $c(i)$ denotes the PSU containing individual $i$. The corresponding
survey weight is
\[
w_i^{(k)}
=
\left\{\pi_i^{(k)}\right\}^{-1}
\approx
\frac{M}{m_k}
\cdot
\frac{\sum_{j\in c(i)}q_{kj}}{n_{k,c(i)}q_{ki}}.
\]

\noindent The internal and external samples are selected independently from the same
finite population using this two-stage procedure, but with different parameter
vectors $\pmb{\gamma}_1$ and $\pmb{\gamma}_2$. This setup creates controlled
differences in the informative sampling mechanisms of $s_1$ and $s_2$, while
preserving the common finite-population target. It allows us to evaluate the
performance of the proposed calibration methods when the two probability
samples have different sampling designs and different weighted compositions
before calibration.

\section{Supplementary Diagnostics for the Data Application}
\label{sec:supp-data-app}

\subsection{Diagnostic for the NHANES 1999--2000 Cycle}
\label{subsec:supp-nhanes99}

\begin{figure}[ht]
\centering
\includegraphics[width=0.85\textwidth]{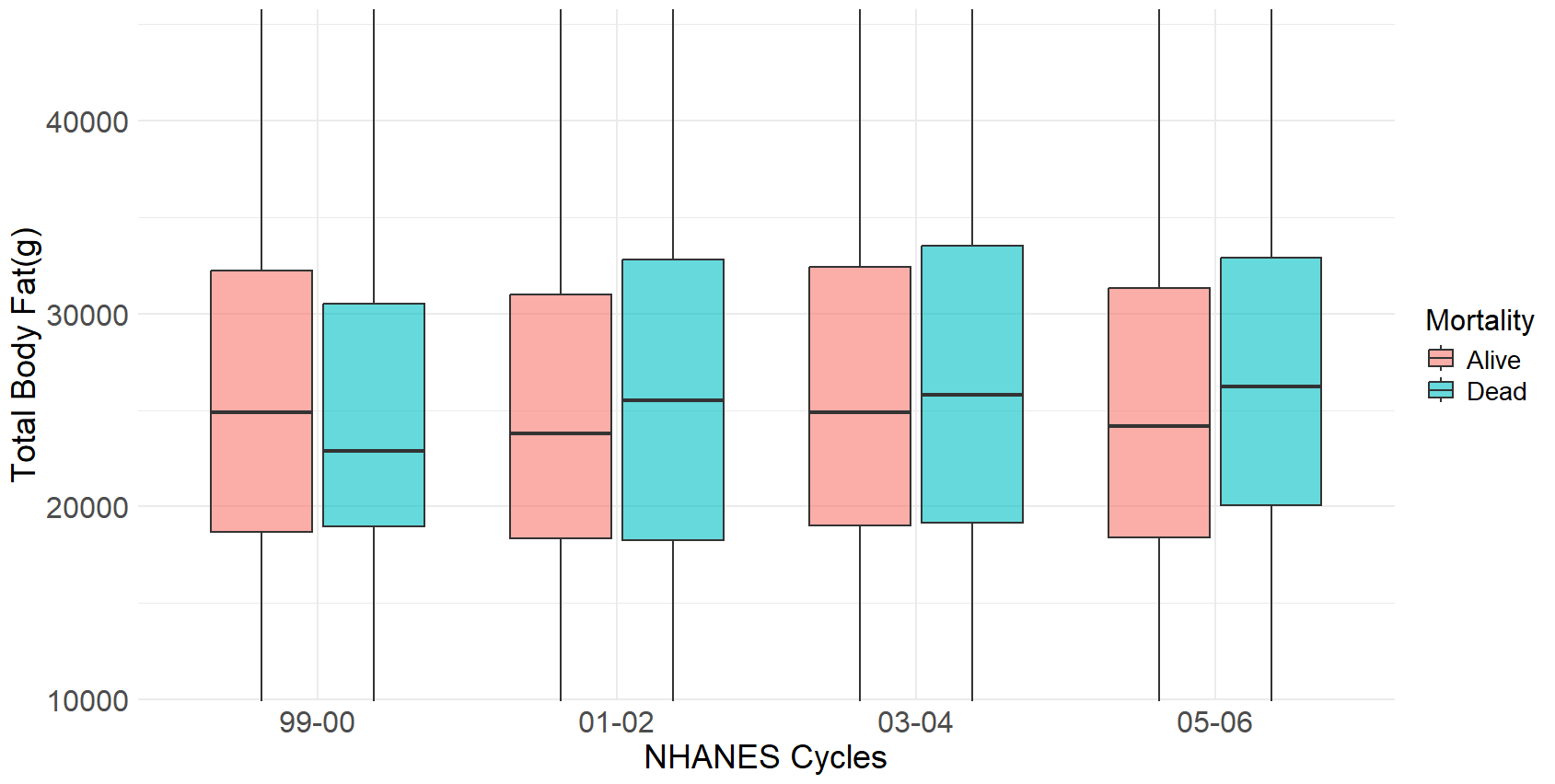}
\caption{Weighted boxplots of total body fat by 10-year all-cause mortality status across NHANES cycles.}
\label{fig:fat_boxplot}
\end{figure}

The calibration framework requires the internal and external samples to be
comparable with respect to the target outcome model, as described in
Assumption~(A.1). As a descriptive check, we examined the distribution of
DXA-measured total body fat across NHANES cycles, stratified by 10-year
all-cause mortality status. Figure~\ref{fig:fat_boxplot} summarizes this
diagnostic.

\begin{figure}[H]
\centering
\includegraphics[width=0.85\textwidth]{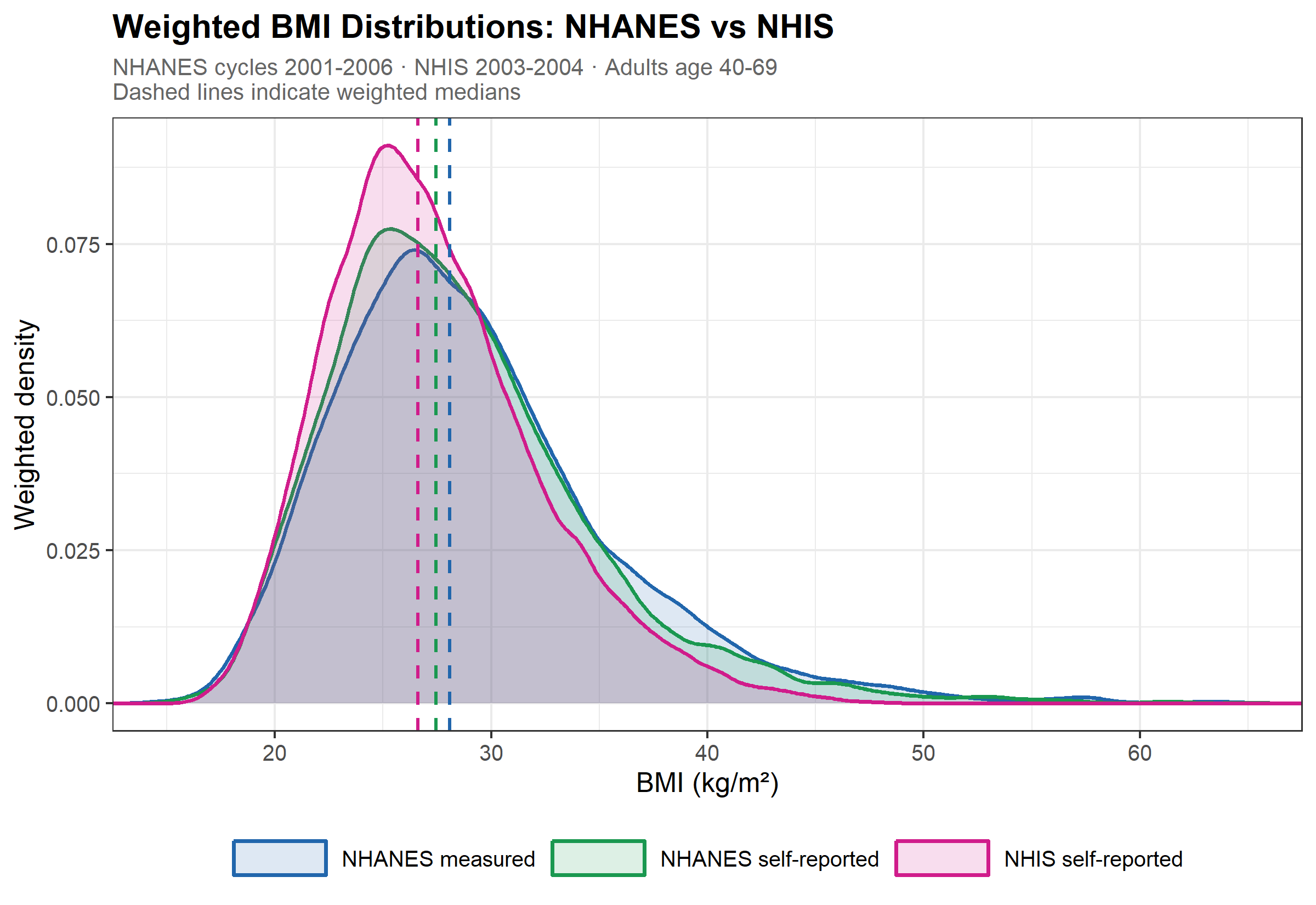}
\caption{Weighted BMI density distributions comparing NHANES measured BMI, NHANES self-reported BMI, and NHIS self-reported BMI among adults aged 40--69. Dashed vertical lines indicate weighted medians.}
\label{fig:bmi_distribution}
\end{figure}

The 1999--2000 cycle appears less similar to the later cycles, particularly
among decedents. This difference may reflect changes in measurement protocols,
eligibility criteria, or cohort composition. To reduce potential
incompatibility across cycles, the main analysis uses the 2001--2002 and
2005--2006 cycles as the NHANES external sample and excludes the 1999--2000
cycle.

\subsection{Diagnostics for the Confidence-Interval Plot}
\label{subsec:supp-ci-diag}

Figure~4 of the main paper presents point estimates and 95\% confidence
intervals for all regression coefficients. Some differences across estimators,
particularly for Hispanic, NH-Other, and smoking-related coefficients, may
reflect differences in sample size, covariate distributions, and
category-specific event counts across data sources. Table~\ref{tab:eda-sample-summary}
provides descriptive summaries after applying the main analysis inclusion
criteria.

\begin{table}[htbp]
\centering
\caption{
Descriptive summaries of the data sources after applying the inclusion criteria
of the main analysis.
}
\label{tab:eda-sample-summary}

\scriptsize
\setlength{\tabcolsep}{3pt}
\renewcommand{\arraystretch}{1.05}

\begin{adjustbox}{max width=\textwidth,center}
\begin{tabular}{lrrrrrrrr}
\toprule
\multicolumn{9}{l}{\textit{Panel A: Sample size, all-cause mortality, and race/ethnicity}} \\
\midrule
Data source
  & $n$
  & \shortstack{Deaths\\$n$ (\%)}
  & \shortstack{Female\\(\%)}
  & \shortstack{NH Black\\(\%)}
  & \shortstack{Hispanic\\$n$ (\%)}
  & \shortstack{Hispanic\\deaths}
  & \shortstack{NH Other\\$n$ (\%)}
  & \shortstack{NH Other\\deaths} \\
\midrule
\textit{NHANES 1999--2000}$^{\dagger}$ & 1487 & 154 (8.4\%) & 50.5 & 8.9 & 544 (11.3\%) & 51 & 39 (4.3\%) & 5 \\
\addlinespace[2pt]
NHANES 2001--2002                   & 1701 & 145 (7.0\%) & 49.9 & 9.6 & 449 (10.5\%) & 29 & 61 (4.5\%) & 3 \\
NHANES 2003--2004 (internal, $s_1$) & 1543 & 146 (6.4\%) & 51.4 & 9.8 & 375 (9.1\%)  & 41 & 65 (4.8\%) & 6 \\
NHANES 2005--2006                   & 1406 & 108 (5.6\%) & 51.2 & 9.2 & 352 (10.0\%) & 23 & 55 (4.5\%) & 3 \\
\addlinespace[2pt]
NHIS 2003--2004 (external, $s_2$)   & 24230 & 2254 (8.5\%) & 49.7 & 10.2 & 3473 (9.2\%) & 237 & 908 (4.2\%) & 69 \\
\bottomrule
\end{tabular}
\end{adjustbox}

\vspace{0.75em}

\begin{adjustbox}{max width=\textwidth,center}
\begin{tabular}{lrrrrrr}
\toprule
\multicolumn{7}{l}{\textit{Panel B: Smoking status and alcohol use}} \\
\midrule
Data source
  & \shortstack{Former Smoker\\$n$ (\%)}
  & \shortstack{Former Smoker\\deaths}
  & \shortstack{Current Smoker\\$n$ (\%)}
  & \shortstack{Current Smoker\\deaths}
  & \shortstack{Some-day\\drinker\\(\%)}
  & \shortstack{Heavy\\drinker\\(\%)} \\
\midrule
\textit{NHANES 1999--2000}$^{\dagger}$ & 457 (31.5\%) & 50 & 325 (22.6\%) & 58 & 44.1 & 9.8 \\
\addlinespace[2pt]
NHANES 2001--2002                   & 497 (29.1\%) & 49 & 404 (22.9\%) & 54 & 45.5 & 8.2 \\
NHANES 2003--2004 (internal, $s_1$) & 451 (30.0\%) & 44 & 381 (24.0\%) & 55 & 42.8 & 8.3 \\
NHANES 2005--2006                   & 400 (29.0\%) & 38 & 352 (24.6\%) & 47 & 44.4 & 10.6 \\
\addlinespace[2pt]
NHIS 2003--2004 (external, $s_2$)   & 6330 (27.2\%) & 659 & 5814 (23.0\%) & 861 & 43.1 & 5.0 \\
\bottomrule
\end{tabular}
\end{adjustbox}

\vspace{0.4em}
\begin{minipage}{0.95\textwidth}
\footnotesize
\textit{Note.}
The inclusion criteria are age $\geq 40$, valid DXA measurement,
valid self-reported BMI $\geq 18.5$, and height between 120 and 220 cm.
Percentages are weighted using the corresponding survey sampling weights.
For selected subgroups, unweighted counts and the number of deaths are also
reported to help interpret the confidence-interval plot in the main paper.
$^{\dagger}$NHANES 1999--2000 is excluded from the main analysis because its
DXA-measured total body fat distribution appears less comparable with later
NHANES cycles; see Figure~\ref{fig:fat_boxplot}.
\end{minipage}
\end{table}

\medskip
The patterns in Figure~4 of the main paper are broadly consistent with the
descriptive summaries in Table~\ref{tab:eda-sample-summary}. Some coefficients
are estimated from relatively small category-specific event counts in the
NHANES samples; for example, the NHANES 2003--2004 internal sample contains
451 former smokers with 44 deaths, and the NH-Other category contains only
55--65 individuals with 3--6 deaths in each NHANES cycle. Such sparse or
moderately sparse cells may contribute to wider confidence intervals in
NHANES-based analyses. By contrast, the NHIS external sample contains many more
individuals in these categories, including 6{,}330 former smokers with 659
deaths and 908 NH-Other participants with 69 deaths, which helps explain the
narrower confidence intervals under NHIS-based calibration. At the same time,
some shifts in NHIS-based point estimates, such as those for Hispanic and
heavy-drinker coefficients, may reflect differences in sample composition,
category-specific mortality patterns, questionnaire wording, or variable
harmonisation across NHANES and NHIS. Thus, the CI patterns likely reflect both
the larger information content of NHIS and modest distributional or measurement
differences between the two surveys.

\subsection{Supplementary Table of Point Estimates and Confidence Intervals}
\label{subsec:supp-ci-table}

The following table reports the numerical values underlying Figure~4 of the
main paper, including point estimates, standard errors, and 95\% confidence
intervals for all regression coefficients and estimators in the data
application. These include the pooled NHANES 2001--2006 benchmark, the
NHANES 2003--2004 internal-sample estimator, and the pooled- and external-
sample calibration estimators using either NHANES 2001--2002 and 2005--2006
or NHIS 2003--2004 as the external sample. Variances for the calibration
estimators are computed using the Taylor-linearized variance formula described
in Section~5 of the main paper and derived in Section~2 of this supplement.

\newpage

\begin{sidewaystable}[p]
\centering
\caption{Point Estimates and 95\% confidence intervals of log-Odds Ratios for 10-year all-cause mortality.}
\label{tab:realdata_balanced}

\footnotesize
\setlength{\tabcolsep}{2.5pt}
\renewcommand{\arraystretch}{1.0}

\begin{adjustbox}{max width=\textheight,center}
\begin{tabular}{lrcrcrcrcrcrc}
\toprule

& \multicolumn{2}{c}{Benchmark}
& \multicolumn{2}{c}{Internal Sample}
& \multicolumn{4}{c}{External Sample NHANES (01-02, 05-06)}
& \multicolumn{4}{c}{External Sample NHIS (03-04)} \\

\cmidrule(lr){2-3}
\cmidrule(lr){4-5}
\cmidrule(lr){6-9}
\cmidrule(lr){10-13}

& \multicolumn{2}{c}{NHANES (01-06)}
& \multicolumn{2}{c}{NHANES (03-04)}
& \multicolumn{2}{c}{Pooled Calibration}
& \multicolumn{2}{c}{External Calibration}
& \multicolumn{2}{c}{Pooled Calibration}
& \multicolumn{2}{c}{External Calibration} \\

\cmidrule(lr){2-3}
\cmidrule(lr){4-5}
\cmidrule(lr){6-7}
\cmidrule(lr){8-9}
\cmidrule(lr){10-11}
\cmidrule(lr){12-13}

Variable
& $\widehat\beta$ & 95\% CI
& $\widehat\beta$ & 95\% CI
& $\widehat\beta$ & 95\% CI
& $\widehat\beta$ & 95\% CI
& $\widehat\beta$ & 95\% CI
& $\widehat\beta$ & 95\% CI \\

\midrule

{\bf Age} ($\times 10^{-2}$) &
9.68  & [8.29, 11.07] &
12.02 & [9.55, 14.49] &
9.70  & [8.34, 11.06] &
8.66  & [6.91, 10.41] &
9.33  & [8.54, 10.13] &
9.21  & [8.47, 9.95] \\
\multicolumn{13}{l}{{\bf Sex} (reference: male)}\\
Female ($\times 10^{-1}$) &
-3.59 & [-6.08, -1.11] &
-6.80 & [-10.98, -2.62] &
-3.61 & [-6.03, -1.19] &
-2.09 & [-5.08, 0.89] &
-5.59 & [-7.01, -4.18] &
-5.53 & [-6.77, -4.30] \\

\multicolumn{13}{l}{{\bf Race/Ethnicity} (reference: Non-Hispanic(NH) White)}\\
NH Black ($\times 10^{-1}$) &
5.87 & [2.87, 8.87] &
4.75 & [0.30, 9.19] &
5.84 & [2.97, 8.70] &
6.65 & [2.65, 10.65] &
5.11 & [3.37, 6.84] &
5.12 & [3.61, 6.64] \\

Hispanic ($\times 10^{-1}$) &
4.39 & [-0.02, 8.80] &
6.73 & [-0.06, 13.52] &
4.39 & [0.27, 9.04] &
3.23 & [-2.44, 8.90] &
-0.66 & [-3.04, 1.71] &
-1.06 & [-3.22, 1.11] \\

NH Other ($\times 10^{-1}$) &
1.83 & [-5.39, 9.05] &
4.30 & [-7.82, 16.41] &
1.84 & [-4.26, 7.95] &
-0.22 & [-9.26, 8.82] &
-2.14 & [-4.67, 0.38] &
-2.55 & [-5.36, 0.27] \\

\multicolumn{13}{l}{{\bf Smoking Status} (reference: non-smoker)}\\
Former smoker ($\times 10^{-1}$) &
7.07 & [3.23, 10.92] &
2.38 & [-3.66, 8.41] &
7.16 & [3.48, 10.84] &
9.61 & [4.65, 14.57] &
3.73 & [2.25, 5.21] &
3.82 & [2.57, 5.07] \\

Current smoker ($\times 10^{-1}$) &
14.51 & [10.75, 18.27] &
13.44 & [7.26, 19.61] &
14.63 & [10.83, 18.42] &
15.54 & [10.65, 20.44] &
11.78 & [9.98, 13.58] &
11.73 & [10.25, 13.20] \\
\multicolumn{13}{l}{{\bf Drinking} (reference: non-drinker)}\\
Some-day drinker ($\times 10^{-1}$) &
-4.19 & [-7.53, -0.85] &
-7.13 & [-12.24, -2.02] &
-4.22 & [-7.34, -1.10] &
-2.89 & [-7.22, 1.44] &
-5.63 & [-6.93, -4.33] &
-5.56 & [-6.68, -4.44] \\

Heavy drinker ($\times 10^{-2}$) &
-2.87 & [-52.19, 46.46] &
-2.01 & [-48.64, 44.61] &
-3.61 & [-53.54, 46.31] &
-3.77 & [-75.46, 67.93] &
1.92  & [-15.74, 19.58] &
2.99  & [-21.67, 27.65] \\

{\bf Physical activity} ($\times 10^{-2}$) &
-4.23 & [-10.61, 2.16] &
-3.36 & [-20.65, 13.94] &
-4.40 & [-12.73, 3.92] &
-4.53 & [-10.69, 1.63] &
-3.78 & [-14.66, 7.10] &
-3.86 & [-11.05, 3.33] \\

{\bf Total fat} ($\times 10^{-6}$) &
13.12 & [-2.19, 28.44] &
13.00 & [-11.69, 37.70] &
13.56 & [-5.38, 32.51] &
14.93 & [-6.37, 36.23] &
14.84 & [1.86, 27.82] &
14.98 & [3.15, 26.82] \\

\bottomrule
\end{tabular}
\end{adjustbox}
\end{sidewaystable}